\documentclass[final,3p,10pt]{elsarticle}

\usepackage[english]{babel}
\usepackage{amsmath,amssymb,amsfonts,amsthm}
\usepackage{makecell}

\usepackage{xcolor}

\biboptions{sort&compress,super}

\usepackage[colorlinks=true,linkcolor=blue,citecolor=red]{hyperref}

\usepackage{listings}
\usepackage{array}
\usepackage{tabularx}
\usepackage{ltablex}

\newcolumntype{T}[1]{>{\tt\footnotesize\raggedright\arraybackslash}p{#1}}
\newcolumntype{D}[1]{>{\it\footnotesize\raggedright\arraybackslash}p{#1}}
\newcolumntype{M}[1]{>{\scriptsize\raggedright\arraybackslash}p{#1}}

\newcommand{\onlinecite}[1]{\nocite{#1}\hspace{-0.1cm}\citenum{#1}}

\newcommand{\figu}[1]
{Fig.~\ref{#1}}

\newcommand{\secu}[1]
{Sec.~\ref{#1}}

\newcommand{\ket}[1]
{|#1\rangle}

\newcommand{\bra}[1]
{\langle #1|}

\def\bcen{\begin{center}}
\def\ecen{\end{center}}

\def\a{\alpha}       \def\b{\beta}      
\def\e{\varepsilon}          
              
                    \def\s{\sigma}

\def\PP{{\cal P}}\def\MM{{\cal M}} 
\def\FF{{\cal F}}\def\HH{{\cal H}}
\def\TT{{\cal T}}\def\BB{{\cal B}} 
 \def\OO{{\cal O}}
\def\DD{{\cal D}}
\def\AA{{\cal A}}
\def\GG{{\cal G}} \def\SS{{\cal S}}

\def\RRR{\mathbb{R}} \def\CCC{\mathbb{C}}

\def\=={\equiv}

\def\qed{\raise1pt\hbox{\vrule height5pt width5pt depth0pt}}

\def\cG0{{\cal G}_0}
\def\cG{{\cal G}}    
  
\def\up{\uparrow}  \def\dw{\downarrow}

\def\=={\equiv}
 
  \def\Tr{{\rm Tr}\,}

\def\ibra{\langle}
\def\iket{\rangle}

\usepackage{bbold}
\def\11{\mathbb{1}}
\def\00{\mathbf{0}}
\def\NAME{{\rm EDIpack}}

\journal{Computer Physics Communications}

\begin{document}

\begin{frontmatter}

\title{\NAME: A parallel exact diagonalization package for
  quantum impurity problems}
\author[a]{A.~Amaricci\corref{author}}
\author[b,c]{L.~Crippa}
\author[b]{A.~Scazzola}
\author[d]{F.~Petocchi}
\author[e]{G.~Mazza}
\author[f]{L.~de Medici}
\author[b]{M.~Capone}

\cortext[author] {Corresponding author.\\\textit{E-mail address:} amaricci@sissa.it}
\address[a]{CNR-IOM, Istituto Officina dei Materiali,
    Consiglio Nazionale delle Ricerche, Via Bonomea 265, 34136 Trieste, Italy}
\address[b]{Scuola Internazionale Superiore di Studi Avanzati (SISSA),
  Via Bonomea 265, 34136 Trieste, Italy}
\address[c]{Institut f\"ur Theoretische Physik und Astrophysik and
  W\"urzburg-Dresden Cluster of Excellence ct.qmat, Universit\"at
  W\"urzburg, 97074 W\"urzburg, Germany}
\address[d]{Department of Physics, University of Fribourg, 1700 Fribourg, Switzerland}
\address[e]{Department of Quantum Matter Physics, University of
  Geneva, Quai Ernest-Ansermet 24, 1211 Geneva, Switzerland}
\address[f]{LPEM, ESPCI Paris, PSL Research University, CNRS, Sorbonne Universit\'e, 75005 Paris, France}

\begin{abstract}
  We present \NAME, an exact diagonalization package to solve
  generic quantum impurity problems. 
  The algorithm includes a generalization of the look-up method
  introduced in Ref.\onlinecite{Lin1993CIP} and enables a massively
  parallel execution of the matrix-vector linear operations required by Lanczos
  and Arnoldi algorithms. We show that a suitable Fock basis organization
  is crucial to optimize the inter-processors communication in a distributed
  memory setup and to reach  sub-linear scaling in sufficiently
  large systems.
  We discuss the algorithm in details indicating how to deal with
  multiple orbitals and electron-phonon coupling. Finally, we outline 
  the download, installation and functioning of the package.
\end{abstract}


\end{frontmatter}

\noindent
{\bf PROGRAM SUMMARY}
\begin{small}
  \noindent
  \\
  {\em Program Title:}  \NAME                                        \\
  {\em CPC Library link to program files:}  \href{https://doi.org/10.17632/2hxhw9zjg9.1}{https://doi.org/10.17632/2hxhw9zjg9.1}                                        \\
  {\em Code Ocean capsule:}  \href{https://codeocean.com/capsule/3537659}{https://codeocean.com/capsule/3537659}                                        \\
{\em Licensing provisions:} GPLv3\\
{\em Programming language:}  Fortran, Python \\
{\em Classification:} 6.5, 7.4, 20 \\
{\em Required dependencies:} CMake ($>=3.0.0$), Scifortran, MPI\\
{\em Nature of problem:} The solution of multi-orbital quantum impurity
systems at zero or low temperatures, including the effective
description of lattice models of strongly correlated electrons, are difficult to determine. \\
{\em Solution method:} Use a parallel exact diagonalization algorithm 
to compute the low lying spectrum and evaluate dynamical correlation functions.\\
\end{small}

\section{Introduction}\label{SecIntro}
The quantum impurity models~\cite{Hewson1993} play a primary role in
the effective description of the local properties of strongly
correlated electron systems~\cite{Kotliar2004PT}.
Indeed, the reduction of complicated lattice models of correlated
electrons to effective impurity systems within dynamical
mean-field
theory~\cite{Georges1996RMP,Kotliar2004PT,Georges2004ACP,Lechermann2006PRB}
(DMFT) or
variational cluster approximations~\cite{Potthoff2003TEPJBCMACS,Senechal2008,Potthoff2011ACP,Nuss2011}
boosted our understanding of correlated materials properties. 
This scientific advance was partially supported by the development of
many numerical methods to solve impurity problems within different approximation schemes~\cite{Bauer2011JOSMTAE,Parcollet2015CPC},
such as the continuous-time quantum Monte
Carlo
approaches~\cite{Gull2011RMP,Rubtsov2005PRB,Haule2007PRB,Seth2016CPC,Wallerberger2019CPC}
the numerical renormalization
 group~\cite{Zitko2009PRB,Bulla2001PRB,Bulla2008RMP} or the
 density-matrix renormalization
 group~\cite{White1992PRL,Hallberg1995PRB,Garcia2004PRL,Schollwock2005RMP,Wolf2015PRX}. 
Among these state-of-the-art methods, the Exact
Diagonalization
(ED) impurity
solver~\cite{Caffarel1994PRL,Dolfen2006,Perroni2007PRB,Capone2007PRB,Weber2012PRB,Lu2017TEPJST}
plays a relevant role.  
The ED method relies on the construction of all or part of the spectrum of a finite
quantum system by solving the associated eigenvalue problem, giving
access to zero or low temperature properties as well as the exact calculation of
one-particle correlation functions on the entire complex plane.
This method is not limited to quantum impurity models and
the ideas presented in this manuscript can be applied also to a wide
range of finite-size quantum
systems~\cite{Weisse2008,Sandvik2010ACP}. 

In the context of DMFT it has been shown that quantum impurity models
with a small Hilbert space already provide accurate results. 
However, the required size rapidly increases if we consider
multi-orbital problems or we include the coupling with phonons.
Likewise, larger systems are also required to resolve small energy or
temperature scales. The exponential growth of the Hilbert space is
the bottleneck of ED calculations. 
In order to overcome such limitations it is mandatory to take advantage of the blossoming
of multi-threaded calculations made available in high-performance
computers. In this respect, it is crucial to route the development of
ED algorithm for impurity problems in the direction of parallel computing~\cite{Siro2012CPC,Borstnik2014PC,Siro2016CPC}.
Yet, the parallel formulation of the ED
algorithms~\cite{Dolfen2006,Siro2016CPC,Siro2012CPC,Sharma2015CPC} can 
be tremendously more complicated with respect to other methods,
e.g. Quantum Monte Carlo, due to the handling of the required 
inter-processor communications.  

  In this work we present the implementation of a 
scalable parallel ED algorithm to solve generic multi-orbital quantum
impurity problems, including the presence of electron-phonon
coupling. 
The goal of our method is to provide an efficient solver for DMFT calculations with arbitrary
electronic band structures at zero or low temperatures, with possible applications to real materials
description through {\it ab-initio}+DMFT
approach~\cite{Kotliar2006RMP}.
The \NAME\ library offers different improvements with respect to the current
implementations available in the literature.
In order to deal with the multi-orbital structure, we introduce a suitable generalization of the look-up
method~\cite{Lin1993CIP,Sharma2015CPC}, originally devised for single orbital systems. 
This generalization takes into account the presence of multiple
quantum numbers  
introducing a great simplification in the construction and
organization of the electronic states.
In addition, we extend this algorithm to the presence of
electron-phonon coupling. 
In the Fock basis, the Hamiltonian is represented by an
exponentially large matrix which makes it impossible to store it, let alone to
entirely diagonalize it. The existence of orbital resolved quantum numbers is
associated to a product structure of the Hamiltonian matrix, which extends the conventional spin resolution
of the hopping terms~\cite{Lin1993CIP,Siro2016CPC,Siro2012CPC,Sharma2015CPC}. 
The organization of the Hilbert space and the construction of a
suitable basis of electronic configurations is a key step in the ED
algorithm.

Exploiting the product structure of the Hamiltonian and its sparse nature, i.e. a
largely reduced number of finite entries with respect to the
total, allows to reduce the memory footprint as well as to employ  
powerful algorithms designed to access part of the
spectrum~\cite{Lanczos1950JRNBSB,ARNOLDI1951QOAM,A.Krylov1931BDLDSDL,Polizzi2009PRB,Sharma2015CPC}.
Such methods generally rely on linear operations, like the
matrix-vector products, which in turn constitute the largest portion
of the computational time.
A generic and efficient optimization of these operations is then critical
to reduce the computational effort.
Here we discuss in details the distributed memory MPI 
algorithms to perform the required matrix-vector operations.
We highlight the bottlenecks created by 
inter-processor communication congestion in a parallel setup
and we show how to unlock a favorable scaling by extending
to the multi-orbital and electron-phonon coupled cases 
the algorithm originally proposed in Ref.\onlinecite{Dolfen2006}.

In particular, we show that under very general conditions the scaling of the ED algorithm is
(sub-)linear in the number of processors, making it possible to solve large systems in a
relatively short time and with a moderate distributed memory use. 
We investigate the effects of multi-orbital interaction on the performances of the parallel
algorithms. While nearly optimal scaling can be achieved including only
density-density interaction terms~\cite{Georges2013ACMP}, the use of a fully
symmetric Kanamori interaction~\cite{Georges2013ACMP} slightly spoils the
scalability of the ED algorithm as a consequence of unavoidable
parallel communication  blockage.
Finally, we highlight that the inclusion of electron-phonon coupling
leads to a linear increase of the total execution time, i.e. it
reduces by a suitable constant factor the favorable parallel scaling.

The rest of this work is organized as follows: In \secu{SecED} we
introduce the generic quantum impurity problem and discuss the
relevant aspects of the ED method  for a generic multi-orbital implementation.
In \secu{SecMVP} we discuss the parallel algorithms designed to
accelerate the execution of the matrix-vector products, at the heart
of the ED method. In \secu{SecEPH} we discuss the extension the ED algorithm to the
case of electron-phonon coupling, including the modifications imposed
to the distributed parallelization. Next, in \secu{SecBenchmarks} we
discuss in details the scaling properties of this algorithm and their
dependence on the number of orbitals and electron-phonon coupling.
In \secu{SecCode} we give an overview of the library structure and
discuss the essential aspects of the code implementation.
Finally, in \secu{SecInstallUse} we 
describe the installation and the use of the library.

\section{Exact Diagonalization}\label{SecED}
\subsection{The multi-orbital quantum impurity problem}\label{sSecQIM}
We consider a system of $N_s$ electronic levels.  A portion $N_\alpha$  of them,
i.e. the impurity levels, interact via a local repulsion, while the
remaining $N_s-N_\alpha$, i.e. the  \textit{bath} levels, are  non-interacting.
In a typical setup the impurity levels are independently coupled to a
set of $N_b$ electronic levels, so that the total number is $N_{s}=N_\alpha 
(N_{b}+1)$. Other choices for the bath topology are possible, which entail a different counting of the total levels.
The Hamiltonian of the electronic system we consider has the form:
\begin{equation}\label{Ham1}
  \begin{split}
    \hat{H}^{e}=&\hat{H}^0+\hat{H}^{int} \\
    \hat{H}^0 = &
    \sum_{\a\b\sigma}
    H^{loc}_{\a\b\sigma}d^{+}_{\a\sigma}d_{\b\sigma}  + \cr
    &\sum_{\nu\a\b\sigma}h^\nu_{\a\b\sigma}a^{+}_{\nu\a\sigma}a_{\nu\b\sigma}+ 
    \sum_{\nu\a\sigma}V^\nu_{\a\sigma}d^{+}_{\a\sigma}a_{\nu\a\sigma}+
    H.c. \\
    \hat{H}^{int}=& U\sum_{\a}n_{\a\uparrow}n_{\a\downarrow}+U'\sum_{\a\neq \b}n_{\a\uparrow}n_{\b\downarrow}+(U'-J)\sum_{\a<\b,\sigma}n_{\a\sigma}n_{\b\sigma}-\\
    &J_X\sum_{\a\neq \b}d^{+}_{\a\uparrow}d_{\a\downarrow}d^{+}_{\b\downarrow}d_{\b\uparrow}+J_P\sum_{\a \neq \b}d^{+}_{\a\uparrow}d^{+}_{\a\downarrow}d_{\b\downarrow}d_{\b\uparrow}
\end{split}
\end{equation}
where $a_{\alpha\sigma}$, $d_{\a\sigma}$ ($a^+_{\alpha\sigma}$,
$d^+_{\a\sigma}$) are, respectively, the destruction (creation) operators for
the bath and impurity electrons with
orbital $\alpha$ and spin $\sigma$,
$n_{\alpha\sigma}=d^{+}_{\alpha\sigma}d_{\alpha\sigma}$. 
The term $H_{\a\b\s}^{loc}$ is the non-interacting part of
the impurity Hamiltonian, 
$h_{\a\b\s}^\nu$ and $V_{\a\s}^\nu$ are, respectively, the local Hamiltonian
and the impurity hybridization of the $\nu-$th bath level. 
  Finally, $\hat{H}^{int}$ is the local
  multi-orbital interaction~\cite{Georges2013ACMP}.
  The first three terms represent the density-density part of the
  interaction, while the remaining two are,
  respectively, the spin-exchange and pair-hopping terms. 
  $U$ is the local Coulomb interaction strength, $J$ is the Hund's
  coupling~\cite{Georges2013ACMP}. We introduced
  independent coupling controlling the spin-exchange and pair-hopping terms,
  respectively $J_X$ and $J_P$. The fully symmetric Kanamori
  interaction is obtained setting $U'=U-2J$ and
  $J_X=J_P=J$~\cite{Georges2013ACMP}.

\subsection{The Fock space and conserved quantum numbers}\label{sSecQNs}
A system of $N_s$ electrons is associated to
a Fock space of the form $\FF_e=\bigoplus_{n=0}^{N_s}
S_-\HH_e^{\otimes n}$,  where $\HH_e=\{\ket{0},\ket{\up},\ket{\dw},\ket{\up\dw} \}$ is the local
Hilbert space of the electrons on a single level,
$S_-$ is the anti-symmetrization operator, $\bigoplus$ is the direct sum and
$\bigotimes$ the tensor product.
The total dimension of the Fock space $\FF_e$ is
$\text{dim}(\HH_e)^{N_s}=4^{N_s}= 2^{2N_s}$. 

The solution of the eigenvalue problem for $H_e$ is simplified by taking
into account the existence of conserved quantities.
The Hamiltonian \eqref{Ham1} conserves the total charge
$\hat{N}$ and the spin component $\hat{S}_z$, as long as we assume
that no terms  breaking the particle number or spin conservation are present,
e.g. local spin-orbital coupling, in-plane magnetic field,
superconducting order, etc.
The conservation of  $\hat{N}$ and $\hat{S}_z$ can be conveniently
re-expressed in terms of conserved number of electrons with spin $\up$ and $\dw$, i.e. $\hat{N}_\up$,
$\hat{N}_\dw$.
Moreover,  if the terms $H^{loc}$ and $h$ are
diagonal and we consider only the density-density part of the
interaction, i.e. $J_X=J_P=0$, the number of electrons with spin $\up$ and $\dw$ is conserved \textit{per
  orbital}, i.e. $\hat{N}^m_\up$, $\hat{N}^m_\dw$, where
$m=1,\dots,N_\alpha$.

In order to formally unify the treatment of these two cases we introduce some convenient
parameters,
  namely  $N_{ud}=1$ $(N_\alpha)$ and
$N_{bit}=N_s$ $(N_s/N_\alpha)$ for conserved total (orbital resolved) number of
electrons with spin $\up$ and $\dw$.
The first corresponds to the number of conserved operators per
spin, the latter indicates the number of electronic levels per spin
(spin and orbital). In the following section we will give a more precise meaning to
$N_{bit}$.

In the rest of this work we will indicate the set of conserved quantum
numbers (QN)s with the tuple $[\vec{N}_\up,\vec{N}_\dw]$, where ($\sigma=\up,\dw$):
\begin{equation}
  \vec{N}_\sigma=
  \begin{cases}
    N_{1\sigma}\==N_\sigma, & \text{if}\ N_{ud}=1 \\
    [N_{1\sigma},\dots,N_{N_\alpha\sigma} ], & \text{if}\ N_{ud}=N_\alpha    
  \end{cases}
\end{equation}
  Note that for $N_{ud}=1$ we identified the single
  component $N_{1\sigma}$ of the vector $\vec{N}_\sigma$ with the
  total number of electrons with spin $\sigma$, i.e. $N_{\sigma}$.

In presence of a set of conserved QNs,  the Fock space decomposes into
a number of sub-spaces of reduced dimension, each
corresponding to a given value of the QNs tuple. We indicate
each sub-space with the term \textit{sector} and the symbol
$\SS[\vec{N}_\up,\vec{N}_\dw]$, or just $\SS$ where no confusion
arises.
The dimension of each sector is given by the number of ways we can
arrange $N_{\a\sigma}$ elements into $N_{bit}$ positions, i.e.:
\begin{equation}
  \begin{split}
  \text{dim}\left(\SS[\vec{N}_\up,\vec{N}_\dw]\right)  & =  \prod_{\a=1}^{N_{ud}}
  \binom{N_{bit}}{N_{\a\up}}
  \prod_{\a=1}^{N_{ud}}
  \binom{N_{bit}}{N_{\a\dw}}\cr
  &\==
  \prod_{\a=1}^{N_{ud}}D_{\a\up}
  \prod_{\a=1}^{N_{ud}}D_{\a\dw}
  \stackrel{\text{def}}{=}D_{\up}D_{\dw} \== D_\SS
  \end{split}
 \label{nud_dimension}
\end{equation}
where we introduced the symbols $D_\sigma$ and $D_{\a\sigma}$  to
indicate, respectively, the dimension of total and orbital spin-subspace 
associated to the given set of QNs. Note that $D_\sigma\==D_{1,\sigma}$ for
$N_{ud}=1$.

\subsection{The  basis states}\label{sSecBasis}
A natural representation of the basis states for the Fock space $\FF_e$
is obtained in the occupation number formalism of second quantization.
The \textit{Fock basis} for a finite system of $N_s$ electrons is
composed of states of the form
$\ket{\vec{n}}=\ket{n_{1\up},\dots,n_{N_s\up},n_{1\dw},\dots,n_{N_s\dw}}$, 
where each element $n_{a\sigma}=0,1$ describes the absence or the
presence of an electron with spin $\sigma$ at the level $a$. In
conjunction with the basis states, we introduce the non-Hermitian, anti-commuting, annihilation and
creation operators $c_{a\sigma}$ and
$c^{+}_{a\sigma}$, respectively. These operators act on the states
$\ket{\vec{n}}$ as:  
\begin{equation}
  \begin{aligned}
    c_{a\sigma}\ket{\vec{n}}=
    \begin{cases}
      (-1)^{\#_{a\sigma}}\ket{\dots,n_{a\sigma}\!-\!1,\dots}
      &\text{if $n_{a\sigma}\!=\! 1$}\\
      0 &\text{otherwise}
    \end{cases};
  \end{aligned}
\end{equation}
    \begin{equation}
  \begin{aligned}
    c^{+}_{a\sigma}\ket{\vec{n}}=
     \begin{cases}
      (-1)^{\#_{a\sigma}}\ket{\dots,n_{a\sigma}\!+\!1,\dots}
      & \text{if $n_{a\sigma}\!=\! 0$}\\
      0 & \text{otherwise}
    \end{cases}    
  \end{aligned}
\end{equation}
with $\#_{a\sigma}=\sum_{b\sigma'<a\sigma} n_{b\sigma'}$. 
Thus, each state is represented as a string of zeros and
ones, i.e. the binary decomposition of a given integer number $I$. Using
the identification $\ket{\vec{n}}=\ket{I}$, with
$I=0,\dots,2^{2N_s}-1$, each state in the Fock space can be encoded in a computer using
a sequence of $2N_s$ bits or, equivalently, an integer number in a
fixed representation.
The exponentially growing size of the Fock space will
eventually make such representation unpractical.
A solution is obtained by decomposing each state according to the
existing QNs. For a given set $[\vec{N}_\up,\vec{N}_\dw]$ of QNs we
then have: 
\begin{equation}
  \begin{split}
  \ket{\vec{n}} &=
  \prod_{\a=1}^{N_{ud}}\prod_{\sigma=\up\dw}\ket{n_1\dots
    n_{N_{bit}}}_{\a\sigma}\cr
  &=
  \begin{cases}
    \ket{\vec{n}_\up}\ket{\vec{n}_\dw},\
    \text{if}\ N_{ud}=1\\
      \ket{\vec{n}_{1\up}}\cdots\ket{\vec{n}_{N_{ud}\up}}
      \ket{\vec{n}_{1\dw}}\cdots\ket{\vec{n}_{N_{ud}\dw}},\
      \text{if}\ N_{ud}=N_\alpha
    \end{cases}
    \end{split}
\end{equation}
so that, if the total number of electrons with spin $\up$ and $\dw$ is 
conserved, any state is identified by two binary sequences of
$N_{bit}=N_s$ bits, one per spin orientation.
Alternatively, if the number of electrons with spin $\up$ and $\dw$ per orbital is conserved, the
states are decomposed into two sets of
binary sequences (one set per spin orientation), each sequence
being made of  $N_{bit}=N_s/N_\alpha$ bits.
Each binary sequence is associated to a suitable tuple
of integer numbers,  univocally identifying the Fock state:
$I \rightarrow [\vec{I}_\up,\vec{I}_\dw]$, 
where $I=0,\dots,2^{2N_s}-1$ and
$\vec{I}_\sigma=[I_{1\sigma},\dots,I_{N_{ud}\sigma}]$ with
$I_{\a\sigma}=0,\dots,2^{N_{bit}}-1$. 
Through such decomposition, each state can be described by the
smallest bit set compatible with the conserved QNs.
This setup generalizes the method introduced by Lin and Gubernatis in
Ref.\onlinecite{Lin1993CIP}. 
The relation
between the Fock state index $I$ and its tuple decomposition can be
easily inverted:
\begin{equation}\label{Irelation}
I = I_1 + \sum_{i=2}^{2N_{ud}}I_i 2^{N_{bit}(i-1)}\,,
\end{equation}
where we rearranged the tuple as $[\vec{I}_\up,\vec{I}_\dw]=[I_1,\dots,I_{2N_{ud}}]$.

Such organization of the Fock states is used to construct
a suitable basis for the sectors $\SS[\vec{N}_\up,\vec{N}_\dw]$.
To any given Fock state $\ket{\vec{n}}$ and its
integer representation $I$, containing the correct bit decomposition
dictated by the QNs, it is associated a state $\ket{i}$ and an integer
$i=1,\dots,D_\SS$ through a suitable \textit{map} $\vec{\MM}_\SS$.
In particular,
each tuple of integers identifying a Fock state belonging to $\SS$ is associated to a new
tuple specific for each sector state:
$I\in\SS=[\vec{I}_\up,\vec{I}_\dw]\in\SS \xrightarrow{\vec{\MM}_\SS}
[\vec{i}_\up,\vec{i}_\dw]=i$, 
where $\vec{i}_\sigma=[i_{1\sigma},\dots,i_{N_{ud}\sigma}]$ and
$i_{\a\sigma}=1,\dots,D_{\a\sigma}$.
The tuple $[\vec{i}_\up,\vec{i}_\dw]$ univocally identifies a basis state
$\ket{i}=\ket{\vec{i}_\up,\vec{i}_\dw}$ of the sector $\SS$, through a
relation similar to \eqref{Irelation}:
\begin{equation}
i = i_{1\up} +
\sum_\sigma\sum_{l=2}^{N_{ud}}(i_{l\sigma}-1)\prod_{\a=1}^{l-1}D_{\a\sigma}
\label{SectorIndxDecomp}
\end{equation}

\subsection{The Hamiltonian matrix}\label{sSecHam}
The matrix representing the system Hamiltonian in the Fock space has a
block-diagonal structure in presence of a given set of conserved QNs.
Each block corresponds to the Hamiltonian of a sector
$\SS[\vec{N}_\up,\vec{N}_\dw]$. The analysis of the spectrum, thus,
reduces to the recursive analysis of the sector Hamiltonians. 
In each sector, the Hamiltonian of the system has the following general form: 
\begin{equation}
H^e_\SS = H_d  + H_\up\otimes \mathbb{I}_\dw + \mathbb{I}_\up\otimes
H_\dw + H_{nd}\,.
\label{Hdecomp}
\end{equation}
$H_d$ is a diagonal term containing the local part of the Hamiltonian,
including the density-density terms of the interaction.
The $H_\sigma$ components describe all the hopping processes of the
electrons with spin $\sigma=\up,\dw$.
Finally the term $H_{nd}$ contains all the remaining non-diagonal
elements which do not fit in the previous two components,
e.g. the spin-exchange and pair-hopping interaction terms.
If the QNs are conserved per orbital, i.e. if $H_{nd}\==0$ and no
inter-orbital local hopping terms are present,  each $H_\sigma$
further splits into a sum of smaller terms:
$$
H_\sigma = \sum_{\a=1}^{N_\alpha}\mathbb{I}_{1\sigma}\otimes\cdots\otimes
H_{\a\sigma}\otimes\cdots\otimes \mathbb{I}_{N_\alpha\sigma}
$$

Each term of the  Hamiltonian matrix is constructed  independently 
iterating over one or more  components of the sector basis states.
For example, the construction of the matrices $H_{\a\sigma}$ only involve iterations over the components
$\ket{i_{\a\sigma}}$ of the sector basis. In general we have:
\begin{equation}
\begin{split}
  \bra{\vec{i}_\up\vec{i}_\dw}H^e_\SS\ket{\vec{j}_\up\vec{j}_\dw}
  =&
  \bra{\vec{i}_\up\vec{i}_\dw}
  H_d + H_\up\otimes \mathbb{I}_\dw + \mathbb{I}_\up\otimes H_\dw + H_{nd}
  \ket{\vec{j}_\up\vec{j}_\dw}\cr
  =&
  \bra{\vec{i}_\up\vec{i}_\dw}H_d\ket{\vec{i}_\up\vec{i}_\dw} +\cr
  &\bra{\vec{i}_\up}H_\up\ket{\vec{j}_\up}\delta_{\vec{i}_\dw\vec{j}_\dw}
  +
  \bra{\vec{i}_\dw}H_\dw\ket{\vec{j}_\dw}\delta_{\vec{i}_\up\vec{j}_\up} + \cr
  &\bra{\vec{i}_\up\vec{i}_\dw}H_{nd}\ket{\vec{j}_\up\vec{j}_\dw} 
\end{split}
\label{Hbuild}
\end{equation}

For very large systems, storing the Hamiltonian matrix in the memory
can be highly inefficient.
In such cases, Krylov or Lanczos
methods~\cite{Lanczos1950JRNBSB,Lin1993CIP,Lehoucq1998,Maschhoff1996} can
be implemented using a storage-free algorithm, performing the
necessary linear operations on-the-fly.
This solution has generally a negative impact on the execution
time, however this can be well compensated by scaling in a massively
parallel framework.

\section{Matrix-vector product:  parallel algorithms}\label{SecMVP}
The bottleneck of any Lanczos based diagonalization is the execution of
matrix-vector product (MVP) operations, which often takes up to 
90\% of the execution time.
For any given sector $\SS$ the MVP of the Hamiltonian with an
arbitrary vector is $\ket{w}=H^e_\SS\ket{v}$ or,
projecting onto the sector basis, $w_i =\sum_{j}[H^e_\SS]_{ij}v_j$.
Distributing the burden of the MVP across multiple processors can
dramatically improve ED calculations and unlock access to larger
systems.
Yet, the MVP heavily relies on communication among different
processors, making the design of efficient parallel algorithms
a non-trivial task. In the following,
we discuss algorithms based on the distributed memory 
\textit{Message Passing Interface }(MPI) framework.

\subsection{The \texttt{MPI\_AllGatherV} algorithm}\label{sSecMPIGather}
A simple and generic parallel algorithm is constructed as follows.
Let us consider the expansion of the vector $\ket{v}$ onto the sector
basis $\ket{v} = \sum_i v_i\ket{i}$.  The vector elements $v_i$ are distributed across the
\texttt{p} processes in data chunks of \texttt{Q = $D_\SS$/p + R} length,
with \texttt{R=mod($D_\SS$,p)} for the first CPU and \texttt{R=0}
otherwise. 
In order to distribute the MVP operation among the processes,
we assign a share of \texttt{Q}  rows of the matrix $H^e_\SS$
to each process, consistently with the splitting of the vector
$\ket{v}$.
In doing that, it is important to distinguish between \textit{local}
and \textit{non-local} elements of the vector, i.e. the share of the
vector which resides in the memory of a given process and the
remaining parts which live on the memory of the other processes.

As such, the result of the MVP consists of two contributions.
The first comes from the product of the $\mathtt{Q}\times\mathtt{Q}$
diagonal blocks of $H^e_\SS$ and the local parts of the
vector $\ket{v}$. This term of the product is executed locally on
each process, i.e. without any inter-process communication. 
The second contribution comes from the multiplication of the
remaining elements of the matrix share and the 
vector. Due to its non-local nature, this term of the MVP requires to
reconstruct the distributed vector $\ket{v}$. This task is achieved through a call to the MPI
function \texttt{MPI\_AllGatherV}, which executes a communication of the
vector shares from each process to all the other processes.

Although conceptually simple, the scaling analysis of  this algorithm
reveals a quick saturation already for a number \texttt{p} of
processors of the order of ten.   
Indeed, the gain deriving
from the decreased size \texttt{Q} of the vector share ($\simeq
1/\texttt{p}$) is rapidly balanced and overcome by the
inter-processors communication in \texttt{MPI\_AllGatherV}, which
amount to a massive data transfer among processors. 
Such communication congestion ultimately prevents to
achieve a good parallel scaling.

\subsection{The \texttt{MPI\_AllToAllV} algorithm}\label{sSecMPIall2all}
An improved, yet less generic, parallel algorithm can be devised 
exploiting the basis states decomposition introduced above~\cite{Dolfen2006}. 
In the sector basis $\ket{\vec{i}_\up\vec{i}_\dw}$ a generic vector $\ket{v}$ can be decomposed as:
\begin{equation}
  \ket{v}
  =
  \sum_i v_i \ket{i}
  =
  \sum_{\vec{i}_\up\vec{i}_\dw}v_{\vec{i}_\up,\vec{i}_\dw}\ket{\vec{i}_\up\vec{i}_\dw}  
  =
  \sum_{i_\up =1}^{D_\up}\sum_{i_\dw =1}^{D_\dw}
  v_{i_\up,i_\dw}\ket{i_\up}\ket{i_\dw}
  \label{Vdecomp1}
\end{equation}
where in the last equality we used the relation
\eqref{SectorIndxDecomp} to rearrange the tuples of indices
$\vec{i}_\up,\vec{i}_\dw$ and the states
$\ket{\vec{i}_\up\vec{i}_\dw}$ in, respectively, a pair of scalar indices
$i_\up, i_\dw$ and states of the form $\ket{i_\up}\ket{i_\dw}$.  
The sector basis states are ordered so that the spin-$\up$ index runs faster than the spin-$\dw$ one. 
Thus, any  vector $\ket{v}$ can be represented as a matrix whose columns correspond to a set of spin-$\up$ components for a given fixed spin-$\dw$ configuration:
\begin{equation}
 \{ v_i \}_{i=1,\dots,D_\up D_\dw}\doteq
  \left(\begin{matrix}
      v_{1_\up {\color{blue} {1_\dw}}}\\
      \vdots \\
      v_{D_\up {\color{blue} {1_\dw}}} \\
%
      %
      \vdots\\
      v_{1_\up {\color{red} {D_{\dw}}}}\\
      \vdots \\
      v_{D_\up {\color{red} {D_{\dw}}}}\\
\end{matrix}
\right)
=
\left(\begin{matrix}
 v_{1_\up {\color{blue} {1_\dw}}} & \dots & v_{1_\up {\color{red} {D_\dw}}} \\
 v_{2_\up {\color{blue} {1_\dw}}} & \dots &v_{2_\up {\color{red} {D_\dw}}} \\
\dots &   \dots & \dots &  \\
 v_{D_\up {\color{blue} {1_\dw}}}& \dots & v_{D_\up {\color{red} {D_\dw}}} \\
\end{matrix}
\right)
\label{Vmatrix}
\end{equation}

Each vector is distributed across the MPI processes such that a number
$\mathtt{Q}_\dw=D_\dw/\mathtt{p}$ of columns is assigned  to each
core.
If present, the rest $\mathtt{R}_\dw=\mathtt{mod}(D_\dw,\mathtt{p})$ is
reassigned to the first $\mathtt{R}_\dw$ processes. This partitioning  corresponds to
distribute  $\mathtt{Q}=D_\up\times \mathtt{Q}_\dw$ vector elements to each
process. Each term in \eqref{Hbuild} contributing to the MVP
is evaluated separately.

\subsubsection{The $H_d$ term}
The diagonal term $H_d$ of the Hamiltonian is distributed among
processes assigning \texttt{Q} rows, i.e. elements, to each core.
Notably, the multiplication $H_d\ket{v}$ takes place locally in the
memory on each process, i.e. without further MPI
communication.

\subsubsection{The ${H_\up\otimes \mathbb{I}_\dw}$ term}
The contribution of the term $H_\up\otimes
\mathbb{I}_\dw$ to the MVP involves only $\up$-electrons elements, for any
fixed configuration of the spin-$\dw$. In the MVP,
this corresponds to run along the elements of each column of the matrix
\eqref{Vmatrix}.
The proposed MPI decomposition ensures that these elements are stored
contiguously in the memory of each process. 
As such, like the diagonal term $H_d$, this term of the MVP can be
computed locally in the memory of any processor, provided the
(small) Hamiltonian $H_\up$ is known to each process or it is evaluated
on-the-fly.

\begin{figure}
  \begin{center}
    \includegraphics[width=1\linewidth]{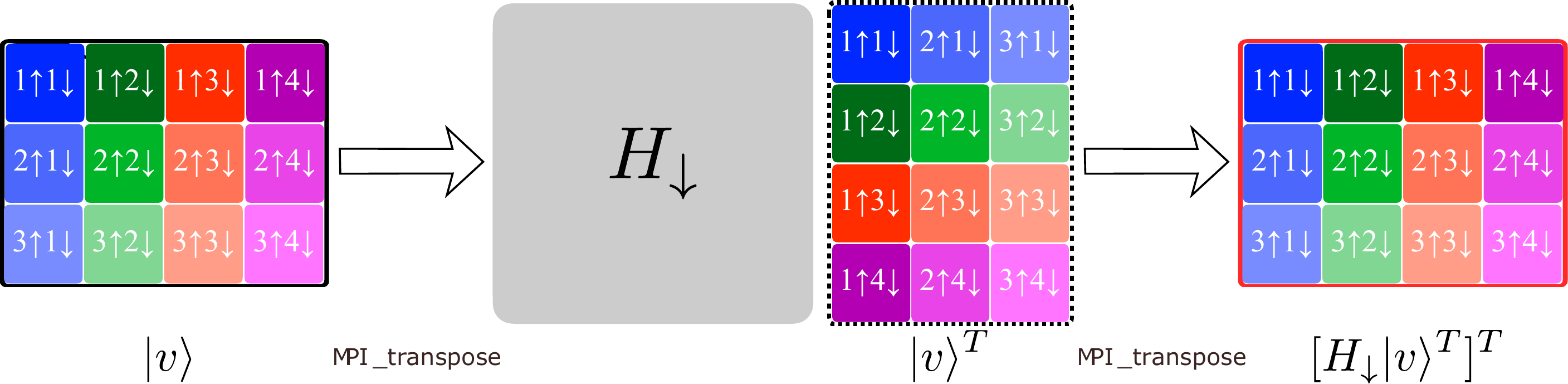}
    \caption{(Color online)
      Schematic representation of the MVP for the
      ${\mathbb{I}_\up\otimes H_\dw}$ term of the sector
      Hamiltonian. In this example the sector dimensions are $D_\SS=D_\up\times
      D_\dw=3\times 4$. We assume to have \texttt{p}=4
      processors. The components of the vector are indicated by their
      index $v_{i_\up i_\dw}\rightarrow i_\up i_\dw$.
      Each column of the initial vector $\ket{v}$  corresponds to a
      different $\dw$ configuration, indicated by different colors.
      The shades of the same color
      correspond to different $\up$ configurations.
      The columns are distributed to the processors. 
      The distributed vector is parallel transposed so that each
      column corresponds to a different $\up$ configuration and belongs to
      one processor.
      The MVP $\ket{w}^T=H_\dw \ket{v}^T$ can then be performed locally in the memory.
      The result $\ket{w}$ is obtained transposing back the resulting
      vector into the original format. 
    }
    \label{Fig1}
  \end{center}
\end{figure}
\subsubsection{The ${\mathbb{I}_\up\otimes H_\dw}$ term}
The product with the term $\mathbb{I}_\up \otimes H_\dw$ involves 
$\dw$-electrons elements, for any fixed spin-$\up$ configuration.
In the MVP this corresponds to run along the rows of the matrix
\eqref{Vmatrix},  ultimately leading to a highly non-local access to
the memory.
In a serial implementation this introduces only a minor performance
degradation due to irregular cache access.
However, in a parallel algorithm such non-locality represents a
serious bottleneck, which requires massive MPI communication
to transfer columns among different processes.
A solution to such problem is to exchange the $\up$-
and $\dw$-configuration indices  $[\vec{i}_\up,\vec{i}_\dw]\rightarrow
[\vec{i}_\dw,\vec{i}_\up]$, corresponding to an actual transposition of the matrix
\eqref{Vmatrix}. The execution of transposed MVP recover locality in the processors
memory.

The key issue, originally pointed out in Ref.~\onlinecite{Dolfen2006},
is reduced to perform a collective transposition of a data set, which
in turn requires a suitable all-to-all communication. 
To this purpose, the MPI library makes available a parallel procedure, 
\texttt{MPI\_AlltoAllv},
which transfers data such that the
\texttt{j}-block, sent from the process \texttt{i}, is
received by process \texttt{j} and placed as block \texttt{i}.
The corresponding communication pattern is schematically depicted in
\figu{figAll2All}.
This \textit{parallel  transposition} involves the minimum amount of
data transfer necessary to execute the MVP, removing the
communicational congestion and unlocking optimal parallel scaling.
Details of the implementation are given in \secu{CodeHamiltonian}.

Summarizing, the MVP of the $\dw$-electrons part of the Hamiltonian proceeds in three steps:
i) the vector $\ket{v}$ is transposed using collective MPI
communication; ii) the multiplication is performed locally on each
process; iii) the resulting vector is transposed back and added up to
the result, see \figu{Fig1}.

\subsubsection{The ${H_{nd}}$ term}
The last contribution to the MVP is from the 
non-diagonal term $H_{nd}$.
This matrix contains elements which can not be reduced to
any favourable form to perform a parallel MVP. 
As such, the execution of the MVP for this term necessarily relies on the 
{\texttt{MPI\_AllGatherV}}-algorithm, discussed above. 
To this end, the matrix $H_{nd}$ gets row-distributed among all
processes assigning \texttt{Q} rows to each core. 
The inclusion of this term is expected to spoil to some 
extent the scalability of the parallel MVP algorithm.

\section{Electron-phonon coupling}\label{SecEPH}
The quantum impurity model (\ref{Ham1}) can be extended to include
electron-phonon coupling. 
The impurity Hamiltonian takes the form: $\hat{H}=\hat{H}_e+\hat{H}_{e-ph} +\hat{H}_{ph}$, where 
$H_{ph}=\omega_{0} b^+b$,
$H_{e-ph}= \sum_{\a\sigma} g_\a
d^+_{\a\sigma}d_{\a\sigma}(b+b^+)$. 
The presence of additional, bosonic, degrees
of freedom introduces major changes to the ED algorithm. However,
such modifications do not spoil the essential aspects of the parallel
algorithms outlined above.

\subsection{Fock space}\label{sSecFockEPH}
The Fock space should be extended to include the presence of the phonons.
In order to deal with the unbounded dimensions of the local Hilbert space of the
phonons we introduce a cut-off $N_{ph}$ to the number of available
phonons. 
The Fock space of a system of $M_{ph}$ phonons is $\FF_{ph}=\bigoplus_{n=0}^{M_{ph}}
S_+\HH_{ph}^{\otimes n}$, where 
$\HH_{ph}=\{\ket{0},\ket{1},\dots,\ket{N_{ph}}\}$ is the local phonon Hilbert space
and $S_+$ the symmetrization operator.  
The Fock space is given by the tensor product of the electronic
and phononic spaces: $\FF=\FF_e\otimes\FF_{ph}$. In the following we
reduce to the case of a single phonon mode $M_{ph}=1$, localized in real space at
the impurity site. As such the dimension of the phonon Fock space is
given by $D_{ph}=N_{ph}+1$.  

The coupling to phonons does
not break the electronic QNs conservation nor it adds novel
symmetries. Thus, the Fock space factorizes into novel sectors given
by the product of any electronic sector and the phonons Hilbert
space. Each sector is then identified by the tuple
$[\vec{N}_\up,\vec{N}_\dw]$ and has dimensions
$\mathrm{dim}\left(\SS[\vec{N}_\up,\vec{N}_\dw]\right)
=D_\up D_\dw D_{ph}\==D_\SS$.

\subsection{Basis states}\label{sSecBasisEPH}
According to this construction, the Fock basis of the electron-phonon
system is composed of product states of the form
$\ket{\vec{n},p}=\ket{n_{1\up},\dots,n_{N_s\up},n_{1\dw},\dots,n_{N_s\dw}}\ket{p}$,
with $p=0,1,\dots,N_{ph}$.  The electronic and phononic creation (annihilation)
operators, respectively, $d^+_{a\s}$, $a^+_{a\s}$ and $b^+$
($d_{a\s}$, $a^+_{a\s}$ and $b$) act
separately on the electronic and phononic part of the states.
This separation of the Fock basis ensures that all the observations 
concerning the electronic configurations hold 
unaltered. In particular, each state $\ket{\vec{n}}$ is repeated
$D_{ph}$ times. Consequently, we identify each Fock state by means of
an integer index and an extended tuple
$I\rightarrow[I_{e},I_{ph}]=[\vec{I}_\up,\vec{I}_\dw,I_{ph}]$, where 
$I_{ph}=0,\dots,N_{ph}$ and $I_e$ identify the electronic Fock state.
The relation between the Fock state index $I$ and the tuple reads:
\begin{equation}
I=I_1+\sum^{2N_{ud}}_{i=2}I_i2^{(i-1)N_{bit}}+I_{ph}2^{2N_s}
\end{equation}

In complete analogy, we use of the map $\vec{\MM}_\SS$ to build a
basis for sector $\SS[\vec{N}_\up,\vec{N}_\dw]$, such that 
$I\in\SS=[\vec{I}_\up,\vec{I}_\dw,I_{ph}]\in\SS \xrightarrow{\vec{\MM}_\SS}
[\vec{i}_\up,\vec{i}_\dw,i_{ph}]=i$, 
where $i_{\a\sigma}=1,\dots,D_{\a\sigma}$ and
$i_{ph}=1,\dots,D_{ph}$. 
The tuple $[\vec{i}_\up,\vec{i}_\dw,i_{ph}]$  identifies a  state
$\ket{i}=\ket{\vec{i}_\up,\vec{i}_\dw}\ket{i_{ph}}$ of the sector
$\SS$ through the  relation:
\begin{equation}
i=i_{1\up}+\sum_{\sigma}\sum_{\alpha=2}^{N_{ud}}(i_{\alpha\sigma}-1)\prod_{\beta=1}^{\alpha-1}
D_{\beta\sigma}+(i_{ph}-1)D_{\uparrow}D_{\downarrow}\,.
\end{equation}

\subsection{Hamiltonian construction}\label{sSecHamEPH}
In each sector $\SS$ the Hamiltonian matrix of the electron-phonon
coupled system has the following expression:
\begin{equation}
  H_\SS= \mathbb{I}^{ph}\otimes H^e_\SS + H^{ph}\otimes \mathbb{I}^{el} + H^{ph}_{\mathrm{e-ph}}\otimes H^{e}_{\mathrm{e-ph}}
  \end{equation}
where $ H^e_\SS$ is the electronic sector Hamiltonian, $H^{ph}$ is the
phonon Hamiltonian, $H^{ph}_{\mathrm{e-ph}}$ and 
$H^{e}_{\mathrm{e-ph}}$ are, respectively, the  phononic and
electronic parts of the electron-phonon coupling term.
Because of the factorization of the basis states, the construction of
the electronic part of the Hamiltonian proceed as outlined above. The
$H^{ph}$ term is diagonal in the phonon basis and independent of the
electronic configuration. The electron-phonon coupling term
$H^{ph}_{\mathrm{e-ph}}\otimes H^{e}_{\mathrm{e-ph}}$ can be
factorized. The electronic part is  diagonal, as the component $H^e_{\mathrm{e-ph}}$ is
proportional to the electronic density operators. However, the phonon
factor $H^{ph}_{\mathrm{e-ph}}$, being linear in the phonon displacement
operator $\hat{x}=(b^+ + b)$, is off-diagonal in the phonon states.

\begin{equation}
\begin{split}
  \bra{\vec{i}_\up\vec{i}_\dw}\bra{i_{ph}}H_\SS\ket{\vec{j}_\up\vec{j}_\dw}\ket{j_{ph}}
  =&
  \bra{\vec{i}_\up\vec{i}_\dw}H^e_\SS\ket{\vec{j}_\up\vec{j}_\dw}\delta_{i_{ph}j_{ph}}
  +\cr
  &\delta_{\vec{i}_\up\vec{j}_\up}\delta_{\vec{i}_\dw\vec{j}_\dw}
  \bra{i_{ph}}H^{ph}\ket{i_{ph}} + \cr  
  &
  \bra{\vec{i}_\up\vec{i}_\dw} H^{e}_{\mathrm{e-ph}} \ket{\vec{i}_\up\vec{i}_\dw}
  \bra{i_{ph}} H^{ph}_{\mathrm{e-ph}} \ket{j_{ph}}\cr
\end{split}
\label{HephBuild}
\end{equation}

\begin{figure}
  \begin{center}
    \includegraphics[width=0.45\linewidth]{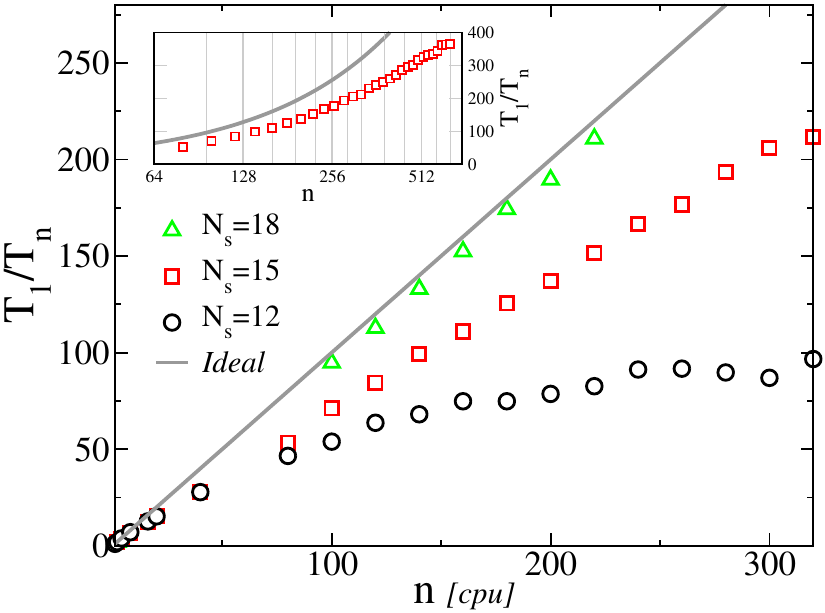}
    \includegraphics[width=0.45\linewidth]{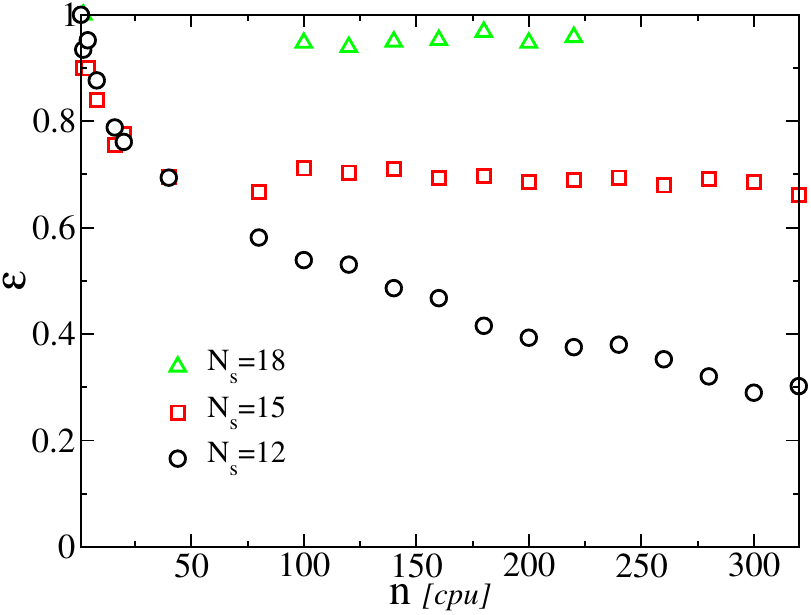}
        \caption{(Color online)
          Left: Parallel speed-up $T_1/T_n$ as a function
          of the CPUs number $n$. Data are for systems with increasing number of
          levels $N_{s}=12$, $15$ and $18$. The ideal scaling (solid gray
          line) is reported for comparison.
          The parallel gain saturates at $n\simeq 256$ for $N_s=12$, it shows
          sub-ideal behavior for $N_s=15$ and  nearly ideal scaling for
          $N_s=18$. Data for $N_s=18$ are normalized to $n=2$. 
          Inset: Parallel speed-up as a function of CPUs number for
          $N_s=15$ in a wider range of $n$ and on a semi-log scale. The results
          shows the saturation of the parallel gain for $n> 512$.
          Right: Parallel efficiency $\e=T_1/nT_n$ as a function  of
          the CPUs number $n$. Data as in left panel. 
        }
        \label{fig2}
  \end{center}
\end{figure}

\subsection{Parallel matrix-vector product}\label{sSecMVPEPH}
The factorization of the electrons and phonons configurations in the
basis states allows to introduce only few changes to the MVP
algorithm, designed to optimize the parallel execution.
The most important step concerns the distribution of the vectors among
the processes. To this end, we observe that the vector decomposition
along the sector basis is such that the electronic configurations are
``repeated'' $D_{ph}$ number of times:
\begin{equation}
  \ket{v}
  =
  \sum_{i,i_{ph}} v_{i,i_{ph}} \ket{i}\ket{i_{ph}}
  =
  \sum_{i_\up =1}^{D_\up}\sum_{i_\dw =1}^{D_\dw}\sum_{i_{ph}=1}^{D_{ph}}
  v_{i_\up,i_\dw,i_{ph}}
  \ket{i_\up}\ket{i_\dw}\ket{i_{ph}}
  \label{Vdecomp2}
\end{equation}
Thus, each vector is distributed  among the MPI processes
such that $\mathtt{Q}^{ph}_\dw \times D_{ph}=D_\dw/\mathtt{p}\times
D_{ph}$ columns, i.e. $D_{ph}$ copies of
the $\mathtt{Q}_\dw$ electronic configurations, are assigned to each
core. Each process holds  $\mathtt{Q}^{ph}=D_\up\times
\mathtt{Q}_\dw\times D_{ph}\==D_\up\times\mathtt{Q}^{ph}_\dw$  elements. 

Then, for any fixed phonon configuration, the execution of the MVP
for the purely electronic part of the
Hamiltonian proceeds along the same lines.    
The product with the phonon Hamiltonian $H^{ph}$,  diagonal in the electronic states, 
only involves iterations along the phonon index for all the electronic
elements residing on each processor, i.e. it is local in the memory.
Finally, according to the vector distribution among the processors,
the product with the electron-phonon Hamiltonian is straightforward.
The $H^e_{\mathrm{e-ph}}$ Hamiltonian is diagonal and distributed to the
processors in shares of size $\mathtt{Q}$ rows, so its contribution is
local in the memory.
The term $H^{ph}_{\mathrm{e-ph}}$, known to each process, connects
 $\mathtt{Q}$ columns of electronic elements having different phonon
index, yet residing in the memory of the same process and requiring
no further change to the MVP algorithm.

\section{Benchmarks} \label{SecBenchmarks}
We present some benchmark results for the massively
parallel ED algorithm outlined above. The calculations have been performed on
HPC cluster, using Intel Xeon E5-2680 v2 processors with
2 sockets, 10 cores, 2 threads per core and 40 GB RAM.
We considered a multi-orbital quantum impurity model with a total of $N_s$
electronic levels. The impurity hosts $N_\a$ orbitals, each coupled via
an hopping amplitude to $N_b$ bath levels, with random energies in the
interval $[-2D,2D]$, with $D=1$ setting the energy unit.
The tests have been executed using total spin $\up$, $\dw$
electrons occupation QNs in the half-filling sector
$N_\up=N_\dw$.
This setup corresponds to the largest possible sector dimension, which
is ideal to test scaling properties of the algorithm.
Timing is relative to the evaluation of the lowest state of the spectrum, using
Lanczos or P-Arpack~\cite{Maschhoff1996} algorithm with on-the-fly evaluation
of the MVP, i.e. without storage of the Hamiltonian. 
In order to reduce the intrinsic errors, we performed repeated
calculations for different realizations of the bath energy distribution.
The data presented in this section are averaged over $10$ realizations.

\begin{figure}
  \begin{center}
    \includegraphics[width=0.45\linewidth]{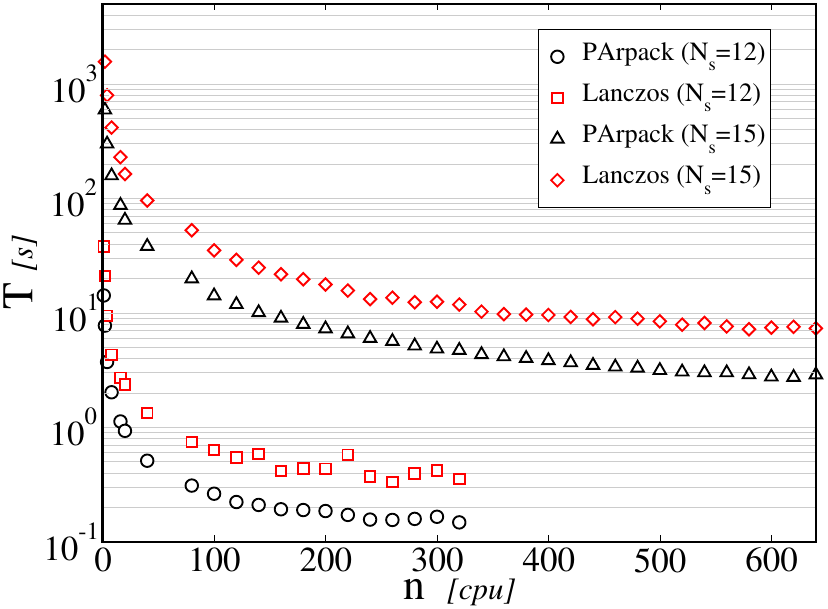}
    \includegraphics[width=0.45\linewidth]{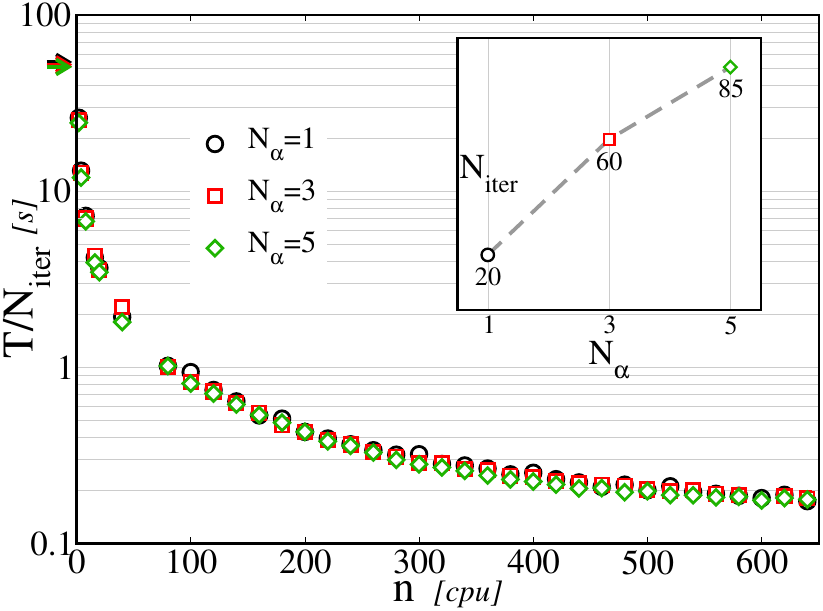}
    \caption{ (Color online)
      Left panel: Total execution time $T$ as a function of the 
      number of CPUs $n$, for systems of $N_{s}=12$, $15$ levels. Data compare
      results obtained with P-Arpack (black circles and triangles) 
      against simple Lanczos (red squares and diamonds)
      methods. 
      Right panel: Average execution time for a single Lanczos
      iteration $T/N_{iter}$ as a function of CPUs number $n$. Data
      are for a system of $N_s=15$ levels with a number of
      orbitals $N_\a=1$, $3$, $5$. The data show that single Lanczos
      iteration time is essentially independent of the orbitals number
      $N_\a$. Inset: number of Lanczos iterations $N_{iter}$ for a single
      groundstate calculation as a function of orbitals number
      $N_\a$. Data show that the iterations number linearly increases with
      $N_\a$ and so does the total execution time $T$.}
    \label{fig3}
  \end{center}
\end{figure}

The speed-up of the parallel algorithm and its efficiency are reported
in  \figu{fig2} for an increasing
number of total electronic levels $N_s$.
For a system with $N_s=12$, the parallel gain rapidly saturates for $n\simeq 250$
CPUs. Indeed, for such small system the ratio between calculation and
communication time quickly becomes unfavourable, leading to a
premature saturation of the performances.
This is further underlined by the decreasing behavior of the parallel
efficiency $\e$ reported in the right panel. 
The scaling improves for larger systems.
In particular, we observed a sub-linear scaling for systems of
$N_s=15$, which displays a saturation tendency only for the largest
accessible number of processes (see inset). For this size, the
parallel efficiency shows a nearly constant behavior up to the largest
number of processor. 
Finally, for the largest system with $N_s=18$ we observed a quasi 
ideal scaling and efficiency near one in the entire range of CPU
number analyzed.

In order to test the solidity of the scaling with respect to the
intrinsic properties of the Lanczos algorithm, we compared the  
single vector Lanczos implementation with respect to a fully fledged  
P-Arpack algorithm~\cite{Maschhoff1996}. For the P-Arpack calculations we
used a block of $\mathtt{ncv}=10$ vectors, which we determined to 
optimize the execution of our program. 
The memory footprint as well as the number of matrix-vector operations
among the two methods are very different, potentially leading
to dissimilar scaling behavior.  
The results in the left panel of Fig.~\ref{fig3} reveal that the 
scaling of the two methods is indeed similar, i.e. the same order of
magnitude. Interestingly however, despite its higher complexity, the more
optimized P-Arpack algorithm performed consistently better than the
plain Lanczos method.

A crucial aspect towards application of the algorithm is the scaling
behaviour with respect to the number of orbitals. 
The multi-orbital interaction  includes terms which are not diagonal
in the electronic configurations, such as spin-flip and
pair-hopping. These terms have a possible detrimental impact on the
parallel scaling, thus we temporarily exclude their contributions and
we will analyze their influence separately.
In multi-orbital systems the hopping matrices 
include more terms with respect to the single orbital, due to the presence of
several hopping channels among orbitals.
In order to characterize quantitatively these aspects we studied the scaling of the
algorithm as a function of an increasing number of orbitals $N_\a$, see
the right panel of  Fig.~\ref{fig3}.
Our results show that the single Lanczos iteration time is essentially
independent of the number of orbitals.
However, the number of iterations required to determine the lowest
eigenstate increases linearly with the number of orbitals, see inset.
We conclude that, although the resulting total time has an excellent 
sub-linear scaling with the number of processes, it linearly depends
on the number of orbitals, i.e. the more the orbitals the larger the
solution time for a fixed number of processes.

\begin{figure}
  \begin{center}
    \includegraphics[width=0.44\linewidth]{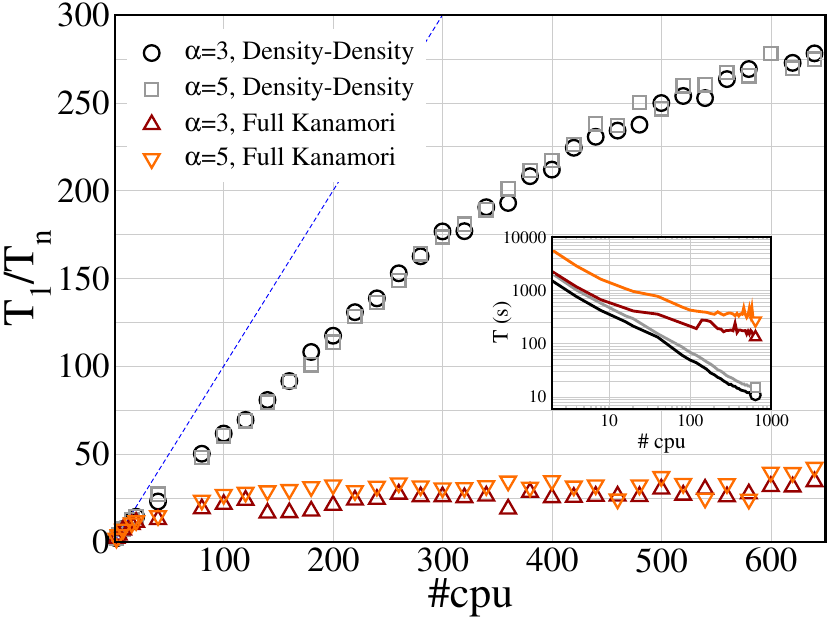}
    \includegraphics[width=0.46\linewidth]{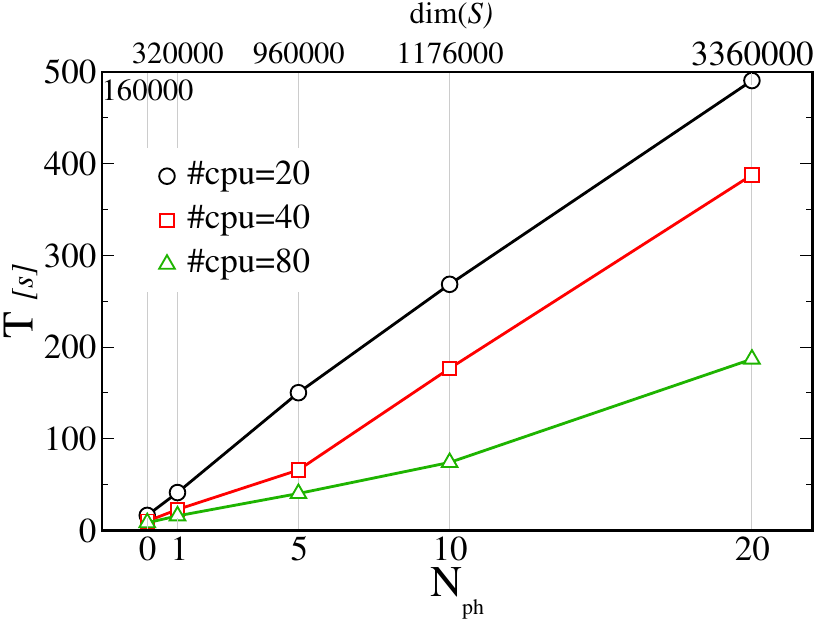}
    \caption{(Color online)
      Left panel: Parallel speedup of the ED algorithm
      as a function of the number of processes for a single
      groundstate calculation. Data are obtained using P-Arpack
      ($ncv=10$), for a number of orbitals $N_\a=3\,,5$, 
      $N_s=15$ and two type of interaction terms: Ising or
      density-density (black circles and gray squares) and full
      Kanamori including spin-flipping and pair-exchange (red up
      triangles and orange down triangles). 
      Ideal scaling (dotted blue line) is reported for comparison. 
      The data show that inclusion of non-diagonal Hamiltonian
      terms $H_{nd}$ quickly saturates the scaling.
      Right panel:
      Total execution time for the groundstate
      calculation as a function of the number of
      phonons $N_{ph}$ (bottom axis) or sector dimension (top
      axis).
      Data are obtained using P-Arpack
      ($ncv=10$), for $N_s=12$ in
      the half filled sector $N_\up=N_\dw=6$ for total occupation
      per spin QNs. The top x-axis shows the dimensions of the
      sector $\mathrm{dim}(\SS)$ for a given number of
      phonons. $N_{ph}=0$ corresponds to the purely electronic problem.            
    }
    \label{fig4}
  \end{center}
\end{figure}

As mentioned above, we analyzed  the effect of non-diagonal
interaction terms on the scaling behavior of our algorithm.
The pair-hopping and spin-exchange terms cannot be reduced to hopping
events involving a single spin orientation. As such we can not take
advantage of the special properties of the
basis states to perform the parallel MVP, possibly spoiling the 
nearly optimal scaling observed for the case of density-density
interaction only.
In order to quantify this effect in \figu{fig4} we compare the
speed-up behavior of the density-density and full Kanamori
interaction, in two systems of $3$ and $5$ orbitals.
As before, the density-density interaction has a sub-linear scaling,
which saturates around $n>600$ processes. The behavior of the full
Kanamori case is however very different.
An initial good speed-up obtained with few tens of processes, quickly
saturates to a value of about $50$ for any number of CPU $n>100$.
In fact, above this threshold the burden of the MPI communication
needed to reconstruct the whole vector and perform the product for
$H_{nd}$ becomes predominant and prevents further improvement of the
scaling. 

The scaling properties of the parallel algorithm are mostly
influenced by the dimensions of the sector associated to a specified
set of QNs as well as the sparseness of the hopping Hamiltonian matrices, which
is partially controlled by the number of orbitals.
Thus, the scaling for the case of orbital resolved QNs is
qualitatively similar to the results discussed above, provided the
sector dimensions are equal, a condition that is fulfilled only for
larger value of  $N_s$.

Finally, we discuss the influence on the scaling of the
electron-phonon coupling. The presence of the phonons leads to a
linear in $N_{ph}$ increase of the total sector dimension, see right panel in  \figu{fig4}.
As the overall cost of the ED solution is proportional to such
dimension, the inclusion of phonons limits the number of electronic
degrees of freedom available to describe the impurity problem.
For instance, with a small number of phonons $N_{ph}\simeq 10$, systems
with up to three orbitals $N_\a=3$ can still be efficiently solved.  
Such an increase in the dimension of the sector caused by
electron-phonon coupling influences the scaling of the parallel MVP
algorithm, leading at most to a linear increase of the time.
Indeed, the product with the electronic part of the Hamiltonian, i.e. the most time
consuming part of the MVP algorithm, should be repeated for each phonon
configuration ultimately reducing by a suitable factor the overall execution performances. 
This expectation is confirmed by the results reported in
the right panel of \figu{fig4}, where we present the scaling of the
ED algorithm with respect to the number of phonons $N_{ph}$ for a system with $N_s=12$ in
the half-filling sector $N_\up=N_\dw$ and total occupation per spin
QNs.
The results show that the total execution time for the
groundstate calculation scales linearly, with a slope that depends
on the CPU number.

\section{Library description and implementation}\label{SecCode}
\begin{figure}
  \begin{center}
    \includegraphics[width=0.9\linewidth]{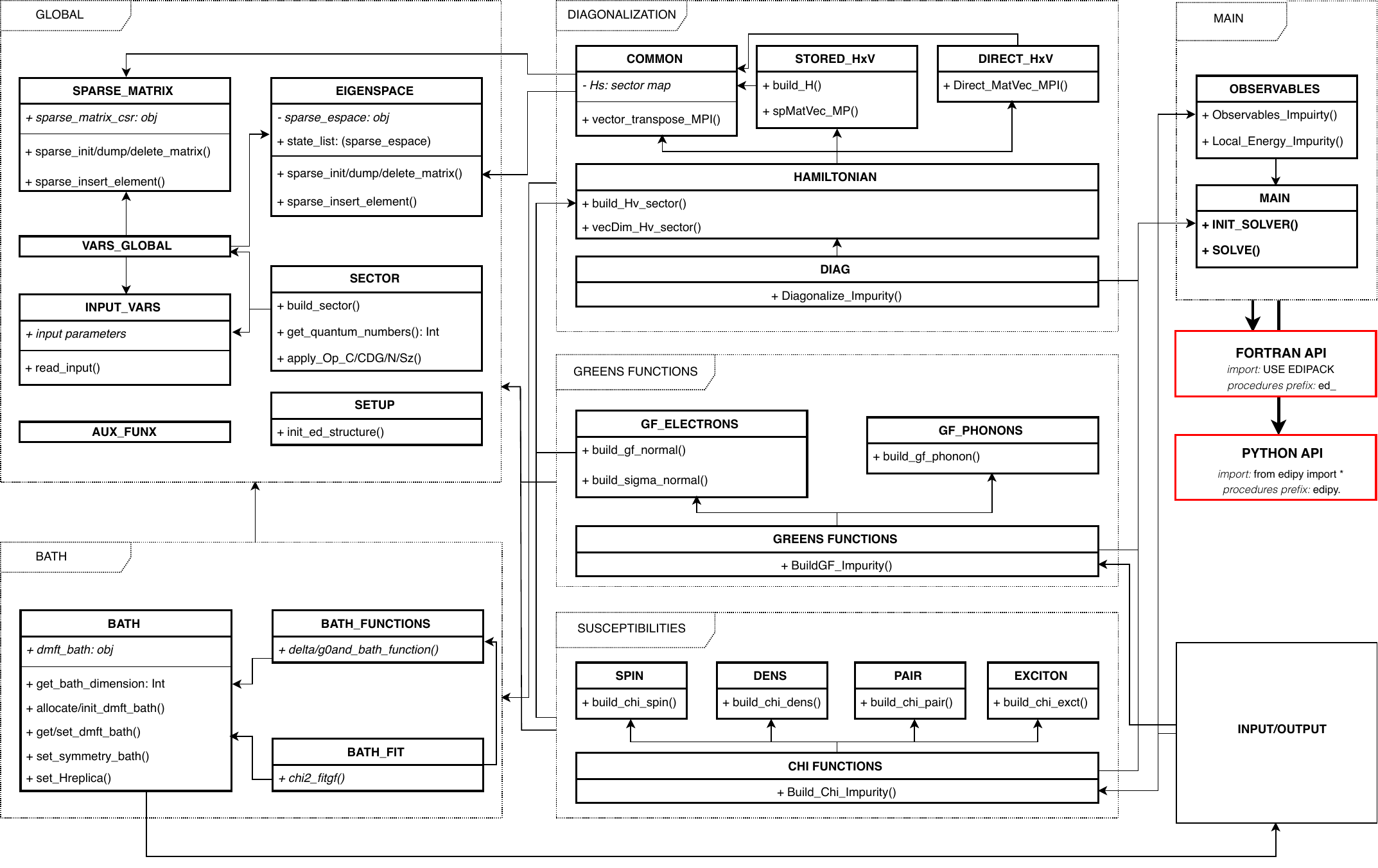}
   	\caption{(Color online) Simplified UML diagram of the \NAME\ 
          library. The library is divided in several abstract work packages
          \textit{Global, Bath, Diagonalization,
            GreensFunctions, Susceptibilities, Input/Output, Main}
          aimed to carry out specific tasks or to contain global
          objects or procedures. 
          General classes and setup procedures are defined in the
          \textit{Global} package. Discretized bath is handled in
          the \textit{Bath} package. The hamiltonian matrix is diagonalized
          in the \textit{Diagonalization} package and its results are
          used in \textit{GreensFunctions}, \textit{Susceptibilities}
          to construct dynamical correlations functions.
          The Fortran API, i.e. the library interfaces, is provided by the top end module
          \texttt{EDIpack}, providing access to a selected
          number of global procedures. 
          The Python binding \texttt{edipy} is built using the \texttt{numpy.f2py}
          tool on a Fortran layer in combination with a pure Python module.  
        } 
   	    \label{figUML}
  \end{center}
\end{figure}
The structure of the  library is schematically represented in
\figu{figUML}, using a simplified Unified Modeling Language (UML)
diagram. The entire library is written in Fortran using object
oriented programming  within a modules structure. 
The Fortran interface is provided by the top-end module exposing to
the end-user the procedures required to setup and solve a quantum
impurity problem.
The Python API are provided by a pure Python module \texttt{edipy}.
  In what follows we describe in more details the structure and the
  implementation of the most relevant parts of the \NAME\ library.  
  In order to unify the notation we
  henceforth indicate with the generic prefix 
  \texttt{?=ed\_} or \texttt{edipy.} either Fortran or Python
  procedures.

\subsection{Sparse matrix class}\label{CodeSparseMatrix}
A sparse matrix storage is performed using a
dedicated custom class, contained in the \texttt{SPARSE\_MATRIX} module. 
The class defines a \texttt{sparse\_matrix\_csr} object as a 
simplified hash-table. The keys corresponds to the rows of the matrix
while the value is associated to a pair dynamical arrays, containing values and
columns location of the non-zero elements of the sparse matrix.
The \texttt{sparse\_matrix\_csr} object can be stored either serially,
i.e. one copy per process, or be parallel distributed assigning a number of keys/values to each
process.
The elements are progressively stored in the dynamic arrays using
\texttt{sp\_insert\_element} procedure, ultimately making use of the Fortran intrinsic
\texttt{move\_alloc}.
This ensures a faster execution compared to implicit  
reallocation, i.e. \texttt{vec=[vec,new\_element]}.
This solution enables to deal with the a priori unknown number of
non-zero elements on each row, to optimize the memory footprint and
to guarantee $O(1)$ access  to any element of the matrix, which are
crucial aspect to speed-up the execution of the MVP. 

\subsection{Eigenspace class}\label{CodeEigenspace}
The storage of the eigenvalues and eigenvectors is a key aspect of
the ED algorithm. In our implementation this task makes use of a
dedicated class, defined in \texttt{EIGENSPACE}.
The class contains the object \texttt{sparse\_espace}: an 
ordered single linked list storing the eigenvalue (the sorting key), the
eigenvector and the corresponding QNs. 
In parallel mode the eigenvectors are automatically distributed to all
processors in shares of size \texttt{Q} (see \secu{SecMVP}) and can be
accessed only
through pointer functions to avoid memory duplication.

For zero temperature calculations only the groundstates are stored
in the list. For a finite temperature also excited states are saved. In
order to avoid unbounded growth of the list we adopt an 
annealing truncation mechanism.
In the first call we collect a number 
\texttt{lanc\_nstates\_sector} of states from each sector, up to a
given maximum number \texttt{lanc\_nstates\_total}. Both parameters
are initially set by the user from the input file.
  The list is truncated to keep  the states which
  fulfil the condition $e^{-\beta(E_i-E_0)} < \mathtt{cutoff}$, where
  $E_i$ is the energy of the $i^{\rm th}$ state in the list, $E_0$ is the
  groundstate energy,  $\beta=1/T$ is the inverse temperature ($k_B=1$) and \texttt{cutoff}
  is an input parameter fixing an a priori energy threshold.
  Annealing is achieved by successive diagonalization of the problem.
  The numbers of states required to any sector $\SS$ contributing to the list is increased by
  \texttt{lanc\_nstates\_step} or it is reduced otherwise. After few calls
  (of the order of ten) the distribution among the sectors of the
  numbers of states reaches a steady state.
  The corresponding annealed list contains all and just the states
  contributing to the spectrum up to the required energy threshold.
  A histogram of the number of states for each sector is produced after
  each diagonalization to check the evolution of their distribution.

\subsection{Sector construction}\label{CodeSector}
The setup of the Fock space sectors is performed in the module  
\texttt{SECTOR}.  
The \texttt{build\_sector} procedure is used to
construct any given sector $\SS[\vec{N}_\up,\vec{N}_\dw]$ by setting
up the map $\vec{\MM}$.
Operationally the map corresponds to a suitable structure which
holds $2N_{ud}$ integer arrays. Any such array has a length
$D_{\a\sigma}$, with $\a=1,\dots,N_{ud}$,
$\sigma=\up,\dw$.  
The arrays are constructed iteratively by looping over the integers
$I_{\a\sigma}=0,\dots,2^{N_{bit}}-1$. The state index is appended into the array if its bit
decomposition corresponds to the required QNs value.
The key part of \texttt{build\_sector} is shown below:
\begin{lstlisting}
 do iud=1,Ns_Ud
    I=0
    do iup=0,2**Nbit-1     
      nup_ = popcnt(iup)  !Intrinsic: count ones in binary decomposition
      if(nup_ /= Nups(iud))cycle
      I  = I+1
      H(iud)%map(I) = iup
    enddo
    I=0
    do idw=0,2**Nbit-1
      ndw_= popcnt(idw)
      if(ndw_ /= Ndws(iud))cycle
      I = I+1
      H(iud+Ns_Ud)%map(I) = idw
    enddo
  enddo
\end{lstlisting}
Inversion of such maps is achieved through a binary search
reverse-lookup algorithm. The small performance loss in access time of
order  $O(N\log(N))$ is overly compensated by the lower memory
footprint.

\begin{figure}
  \begin{center}
    \includegraphics[width=1\linewidth]{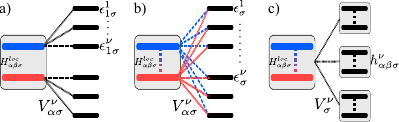}
   	\caption{(Color online) Schematic illustration of the three
          available bath topologies: \texttt{normal} (a),
          \texttt{hybrid} (b) and \texttt{replica} (c).
          (a) In the \texttt{normal} topology a set of $N_b$ electronic levels is
          indipendently coupled
          to each impurity orbital.
          (b) In the \texttt{hybrid} bath a set of $N_b$ of electronic levels is coupled via
          different amplitude to each impurity orbital.
          (c) In the \texttt{replica} bath each electronic level reproduces the structure of the
          impurity and is coupled via a single amplitude. 
        } 
   	    \label{FigBath}
  \end{center}
\end{figure}
\subsection{Bath class}\label{SecBath}
Depending on the symmetries of the quantum impurity problem
different topologies of the non-interacting bath can be devised.
In the \NAME\ library we include three general bath topologies.
The three cases are selected by the input parameter
\texttt{bath\_type=normal, hybrid, replica}~\cite{Koch2008PRB}. A
schematic description of the three cases is reported in
\figu{FigBath}. 

  In the first case, i.e. \texttt{normal}, $N_b$ bath levels are directly coupled to
  each orbital. The total number of electronic levels is then
  $N_s=\mathtt{Norb}(\mathtt{Nbath}+1)$.  
  Each bath level $\nu$ corresponding to the impurity orbital $\a$
bears a local energy
$h^\nu_{\a\b\sigma}=\epsilon^\nu_{\a\sigma}\delta_{\a\b}$
and is coupled to the impurity by an amplitude $V^\nu_{\a\sigma}$.
This bath topology can be used whenever $H^{loc}_{\a\b\sigma}$ is
diagonal in the orbital indices and no particular orbital symmetry breaking
are expected.  Should these conditions not be
verified another bath topology must be considered.

In the \texttt{hybrid} case, the $N_b$ bath levels are all
  coupled to the different impurity orbitals, see \figu{FigBath}b.
  In this case the total number of electronic levels is 
  $N_s=\mathtt{Norb}+\mathtt{Nbath}$.  
  The bath levels carry a local energy
  $h^\nu_{\a\b\sigma}=\epsilon^\nu_{\sigma}$, i.e. independent of the
  orbitals indices, while they hybridize with each impurity orbital with
  amplitude $V^\nu_{\a\sigma}$. In this configuration, the bath
  can describe any effect associated to inter-orbital processes,
  at the cost of introducing possibly redundant bath parameters. 

A way to limit the redundancy of the bath parameters for
  multi-orbital quantum impurity problems with off-diagonal terms is to consider the
  \texttt{replica} bath, see \figu{FigBath}c. In this case the total
  number of electronic levels is 
  $N_s=\mathtt{Norb}(\mathtt{Nbath}+1)$.   
  In this topology each bath level corresponds to a {\it replica} of
  the impurity, i.e. it has the same local structure of the impurity site, with
  parameters to be determined and encoded in the Hamiltonian $h^\nu_{\a\b\sigma}$.
  Each replica couples to the impurity via a single
  amplitude $V^\nu_\sigma$.
  The choice of the structure for $h^\nu_{\a\b\sigma}$ is a key step
  to exploit the nature of this topology.
  To this end, in \NAME\ we assumed that each bath  Hamiltonian can be
  linearly decomposed as
  $$
  h^\nu_{\a\b\sigma}=\sum_{p=1}^{N_p} \lambda^\nu_\sigma(p)
  \Gamma_{\a\b}(p)
  $$
  where the $N_p$ matrices $\Gamma_{\a\b}(p)$ form a suitable basis for the
  local Hamiltonian and $\lambda^\nu_\sigma(p)\in\RRR$  are variational parameters
  for each bath level, i.e. replica. 
  The setup of the matrix basis $\Gamma_{\a\b}(p)$ and the 
  initial values of the variational parameters is possible using the
  interface procedure \texttt{?set\_Hreplica}.  
  This procedure accepts two type of inputs. The first is the local part of the
  impurity Hamiltonian $H^{loc}$, as an array with rank-2 or rank-4
  (see Table \ref{TableProcedures}).
  In this case all the non-zero elements of the $H^{loc}$ are
  identified, their values are used to initialize the paramaters
  $\lambda^\nu_\sigma(p)$. For any
  such non-zero element a matrix $\Gamma(p)$ with the same
  dimension of $H^{loc}$ is defined containing a one only at the 
  position of such element.
  The set of matrices $\Gamma(p)$ naturally decomposes the
  local non-interacting impurity Hamiltonian $H^{loc}$.  
  The second input method requires the user to directly pass the set
  of matrices  $\Gamma(p)$, as rank-5 array and the list of initial
  values for the corresponding variational parameters, see the table
  \ref{TableProcedures}. Although potentially less generic, this
  method allows the user to carefully select the
  basis matrices in order to reflect specific physical aspects, e.g. 
  the existence of possible symmetry breaking.

The bath class is handled using a reverse communication strategy. 
Within the library the bath is described by the class
\texttt{dmft\_bath}. 
On the user side the bath parameters are stored in a rank-1 array of
real numbers, which provides the main input of the solver.
The correct dimensions of such array are retrieved using
\texttt{?get\_bath\_dimension} procedure. 
  Within the \texttt{BATH} class we also provides methods to operate specific
symmetry trasformation onto the bath, e.g. particle-hole, spin
symmetrization, etc. More general operations can be constructed by the
user through suitable methods which allows to get and set  specific
components of the bath.

\subsection{Bath fit}\label{SecBathFit}
  In the DMFT context the bath is updated using a suitable 
optimization of the bath parameters against a given realization of the Weiss field
${\GG^{-1}_{0}}_{\a\b}(z)$, or the corresponding Delta function
$\Delta_{\a\b}(z)= (z + \mu)\delta_{\a\b} - H^{loc}_{\a\b}-
{\GG^{-1}_0}_{\a\b}(z)$, $z\in\CCC$, determined by the self-consistency procedure~\cite{Georges1996RMP}.
This step corresponds to a projection of the Weiss
field (or the Delta function) onto the space of the Anderson quantum
impurity model functions
$\GG^{And}_0(z) =
[(z + \mu)\delta_{\a\b} - H^{loc}_{\a\b} -
\Delta_{\a\b}(z)]^{-1}$, with $\Delta^{And}_{\a\b}(z)=\sum_\nu
{V^\nu_{\a}}(z-h^\nu_{\a\b})^{-1}V^\nu_{\b}$, spanned by the
variational paramaters $\{V,h\}$.

This  optimization can be performed using different
  algorithms~\cite{Garcia2004PRL,Taranto2012PRB}. 
  In \NAME\ we offer to perform this task using a  
conjugate gradient (CG) minimization of the cost function:
$$
\chi = \sum_{n=1}^{L_{fit}}\frac{1}{W_n}\sum_{\a\b}|X_{\a\b}(i\omega_n) - X_{\a\b}^{And}(i\omega_n;\{V,h\})|^q
$$
where $X_{\a\b}={\GG_{0}}_{\a\b},\,\Delta_{\a\b}$ are the user
supplied local functions  and
$X_{\a\b}^{And}={\GG_0}^{And}_{\a\b},\,\Delta^{And}_{\a\b}$ the
Anderson impurity model function, defined above.
The fit procedures are contains in the module \texttt{BATH\_FIT}.  
In order to exploit regularity of the functions the fit is 
performed using imaginary Matsubara frequency. 
The actual form of $X_{\a\b}$ is controlled by the input parameter
\texttt{cg\_Scheme=Weiss,Delta}. 

Given the importance of the fit for the convergence of DMFT calculations, in \NAME\ we
introduced different specific control parameters. This aims to give
the user the possibility of fine tuning the fit and adapt its behavior to
specific situations.
Two distinct CG algorithms can be used in the library, as controlled by the input variable
\texttt{cg\_method=0,1}. The value \texttt{0} corresponds to a
Fletcher-Reeves-Polak-Ribiere minimisation algorithm, adapted from
Numerical Recipes~\cite{NumRec77}. In order to guarantee
back-compatibility with respect to the literature, we also provide the
possibility of using the minimization procedure published in
Ref.\onlinecite{Georges1996RMP} which has been largely used in the DMFT
community.

The gradient $\nabla\chi$ required by the CG algorithm can be
evaluated either analytically or numerically as
controlled by the value of the input parameter \texttt{cg\_grad=0,1}. For
\texttt{cg\_method=1} only numerical gradient can be used.
The number of Matsubara frequency $L_{fit}$ used in the fit procedure
is controlled  by the input parameter \texttt{cg\_Lfit}.
This parameter can be used to restrict the fit to the low frequency
part as the large frequency tails of the local Matsubara functions
have universal behavior.
Similarly, the weight  $W_n=1,1/L_{fit},1/\omega_n$ is used to
enhance the weight of low energy part with respect to the
intermediate-to-large energy one in the cost function.
This is controlled by the input parameter \texttt{cg\_Weight=0,1,2}.
Another  factor contributing to determine the behavior
of the fit is the power $q$ of the cost function $\chi$. By tuning
this parameter, controlled by input variable \texttt{cg\_pow}, it is
possible to enhance or suppress the differences in the $\chi$ in order
to improve (or simplify) the optimization procedure.  

Finally, we introduced few additional parameters which aim to
control the quality of the fit. The first is the fit tolerance,
controlled by the input parameter \texttt{cg\_Ftol}, which
determines the convergence threshold of the minimization procedure.
The second is the maximum number of iterations \texttt{cg\_Niter}
allowed for the CG algorithm to determine the minimum.
The last parameter \texttt{cg\_stop=0,1,2} selects the exit
condition of the minimization procedure. We envisaged  three
possible conditions: $C_1\cup C_2$, $C_1$, $C_2$ with 
$C_1=|\chi^{n-1} -\chi^n|<\mathtt{cg\_Ftol}(1+\chi^n)$ and
$C_2=||x_{n-1} -x_n||<\mathtt{cg\_Ftol}(1+||x_n||)$, where $\chi^n$
is the cost function evaluated at the $n^{\rm th}$-step and $x_n$
the corresponding argument. Increasing the value of the stop
parameter generally loosens up the exit conditions of the minimization.

\subsection{Diagonalization}\label{CodeDiag}
The Hamiltonian diagonalization and the construction of the low-lying
spectrum are performed in the module \texttt{DIAG}.
  For a given set of
conserved QNs, the sector Hamiltonian $H_\SS$ is diagonalized using either
P-Arpack algorithm~\cite{Lehoucq1998,Maschhoff1996} or simple
Lanczos~\cite{Lanczos1950JRNBSB}
procedure with no re-orthogonalization, as controlled by the input variable
\texttt{lanc\_method=arpack,lanczos}.
Sectors with dimensions smaller than the input value \texttt{lanc\_dim\_threshold} are
diagonalized using Lapack method.
The diagonalization requires to visit all the Hamiltonian sectors. If
no long-range order is considered, only half of the sectors can be
considered exploiting a symmetry relation between sectors with
exchanged QNs, i.e.  the eigenvalues of $\SS[\vec{N}_\up,\vec{N}_\dw]$ are identical to those of
$\SS[\vec{N}_\dw,\vec{N}_\up]$, while the corresponding eigenvectors
are identical only up to a trivial reordering. 
The logical variable \texttt{ed\_twin=True,False}  controls this
feature.
In order to further restrict the list of the analysed sectors we allow
the user to indicate explicitly the sectors to solve. This operative mode is
controlled by the variable \texttt{ed\_sectors=True,False}.  
The QNs corresponding to the sector to analyse are read from the file
\texttt{SectorsFile}. Such list can also be partially enlarged by
indicating the number of ``neighboring'' sectors to be solved,
corresponding to QNs shifted by an amount \texttt{ed\_sectors\_shift}
with respect to those on the user list.

\begin{figure}
  \begin{center}
    \includegraphics[width=0.92\linewidth]{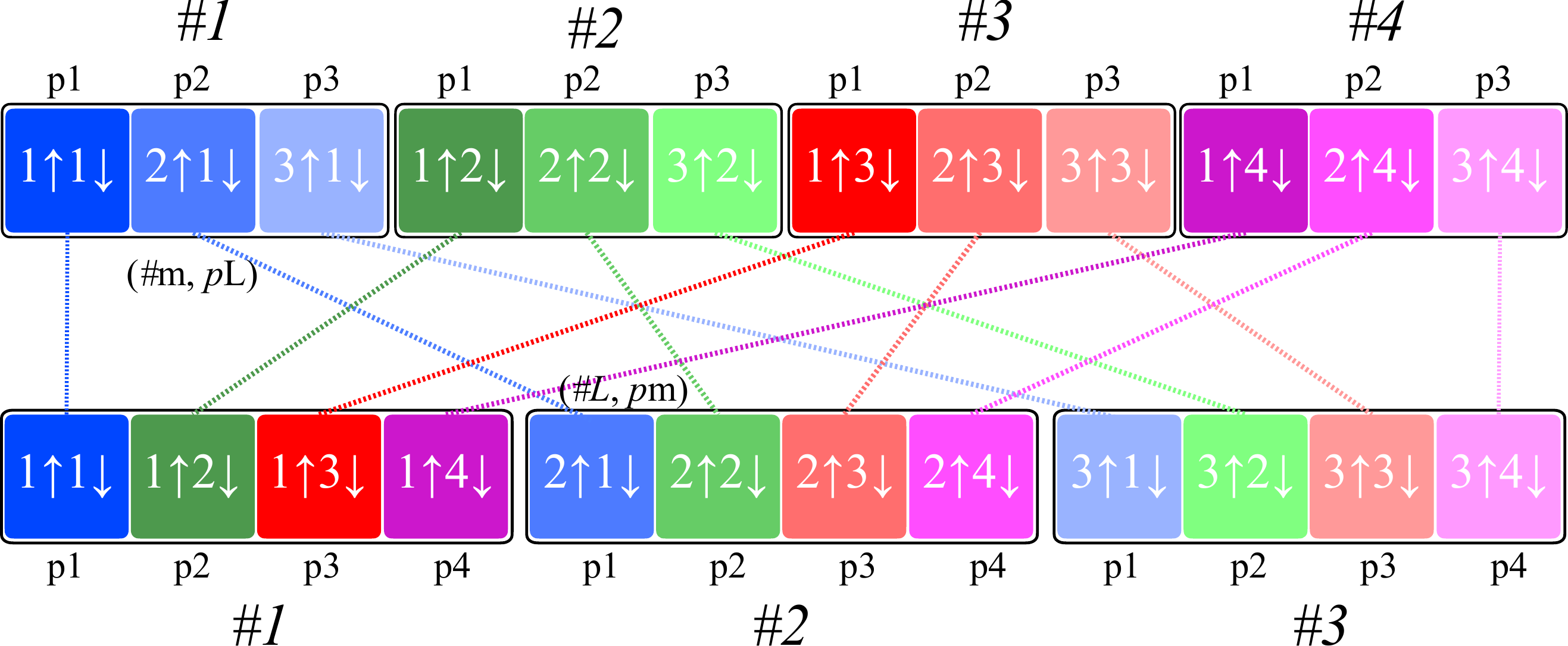}
    \caption{(Color online)
      Schematic representation of the \texttt{MPI\_AllToAllV} parallel
      transposition of a vector. We consider the same case as in
      \figu{Fig1}, i.e. vector dimensions $D_\SS=D_\up\times
      D_\dw=3\times 4$ and \texttt{p}=4 processors.
      The initial vector (top part) is distributed so that each
      processor, indicated with $\#\mathrm{m}$ and $\mathrm{m}=1,\dots,4$, holds one
      column, corresponding to a different color. Each column      
      contains 3 data components, indicated with $p\mathrm{L}$,  
      $\mathrm{L}=1,2,3$ and corresponding to different shades of the
      same color. Each vector component is identified by the
      pair $(\#\mathrm{m},p\mathrm{L})$.
      The procedure \texttt{MPI\_AllToAllV} sends the element at L
      of the processor m to the processor L placing it at
      position m, i.e. each component $(\#\mathrm{m},p\mathrm{L})$ is
      moved to $(\#\mathrm{L},p\mathrm{m})$.
      The final result (bottom part) is the transposed vector.
      Note that processor 4 does not receive any component although it
      participates in the communication process.      
    }
    \label{figAll2All}
  \end{center}
\end{figure}

\subsection{Hamiltonian Construction and Matrix-Vector Product}\label{CodeHamiltonian}
The Hamiltonian matrix construction and the related MVPs operations
are contained in the module  \texttt{HAMILTONIAN}.
Different type of MVP are envisaged depending on the input parameters
\texttt{ed\_Sparse\_H} and \texttt{ed\_total\_ud} which, respectively,
control the storage of the Hamiltonian and the type of QNs.
In order to comply with the different operational modes the MVP is abstracted
into a procedure contained in the global module \texttt{VARS\_GLOBAL}:

\begin{lstlisting}
  abstract interface
    subroutine dd_sparse_HxV(Q,v,Hv)
      integer              :: Q
      real(8),dimension(Q) :: v
      real(8),dimension(Q) :: Hv
    end subroutine dd_sparse_HxV
  end interface
  procedure(dd_sparse_HxV),pointer :: spHtimesV_p=>null()
\end{lstlisting}
The MPI setup, the Hamiltonian construction and the association of the
pointer instance \texttt{spHtimesV\_p} to the correct MVP procedure are
performed in \texttt{build\_Hv\_sector} contained in \texttt{HAMILTONIAN}.
The matrix construction is performed in the module \texttt{HAMILTONIAN\_SPARSE\_HxV}. 
The sector Hamiltonian is represented by a hermitian matrix of dimension $D^2_{\SS}$.
The overall low connectivity of the quantum impurity models reflects
in a strong sparseness  of the Hamiltonian, which hosts at most
$O(N_s)\ll D_\SS$ elements on each row.
We exploit the  Hamiltonian decomposition (\ref{Hdecomp}) to minimize
the memory footprint  storing each matrix component separately. 
The diagonal term $H_{d}$ corresponds to a single vector of length
$D_\SS$. The tensor product structure of the hopping terms can be
readily reconstructed by storing the terms $H_{\a\sigma}$ in a tuple of
sparse matrices,. The total dimension of such tuple is at most of order
$O(N_{ud}N_s\max(D_{\a\sigma}))$,  which is much smaller than $O(N_sD_\SS)$.
Finally, $H_{nd}$ is stored as a separate sparse matrix with a leading
dimension $D_\SS$. This component has a very sparse nature, containing
a number $n\ll N_s$ of elements per row. 

The key part of the \NAME\ library is the parallel execution of the MVPs
using either the memory stored Hamiltonian  or the on-the-fly
execution. All the MVP instances are contained in the two modules
\texttt{HAMILTONIAN\_SPARSE\_HxV}  and
\texttt{HAMILTONIAN\_DIRECT\_HxV}.
As discussed above the fundamental step in the execution of
each MVP is the parallel transposition of the incoming vector.
The corresponding algorithm implementation is presented below: 
\begin{lstlisting}
subroutine vector_transpose_MPI(nrow,qcol,A,ncol,qrow,B)    
    integer  :: nrow,qcol
    integer  :: ncol,qrow
    real(8)  :: A(nrow,qcol)
    real(8)  :: B(ncol,qrow)
    ...
    !A is the input vector of dim(Nrow*Qcol)
    ! on input this is automatically reshaped into a matrix.  
    !Nrow,Ncol: are the global dimensions
    !Qrow,Qcol: are the local dimensions on each processor,
    !B is the transposed vector of dim(Qrow*Ncol)
    ! on output this is automatically reshaped into a vector. 
    !See lines 35 and 54 in the code listing for spMatVec_mpi_main.  
    ... 
    !Get the number Ntot of columns  each processor has to
    !deal with during the AllToAll communication.
    Ntot   = Ncol/MpiSize
    if(mod(Ncol,MpiSize)/=0)Ntot=Ntot+1
    !
    !Build the send_counts buffer for each processor and column
    !held by the processor. It contains the information about how many 
    !elements per column should be sent from CPU irank. 
    counts = Nrow/MpiSize
    do i=1,qcol
       do irank=0,MpiSize-1
          send_counts(irank,i) = counts
          if(irank < mod(Nrow,MpiSize))send_counts(irank,i) = counts+1
       enddo
    enddo
    !
    !The corresponding receive buffer recv_counts is built 
    !by transposition of the send buffer, performed using MPI_AllToAll.
    do i=1,Ntot
       call MPI_AllToAll(send_counts(0:,i),1,MPI_INTEGER,recv_counts(0:,i),1,&
                         MPI_INTEGER,MpiComm,ierr)
    enddo
    !    
    !Next: get the position in memory of the elements. This is build  
    !as a displacement with respect to starting memory addresses. 
    !So this amount to count how many elements have been communicated  
    !from previous processors.    
    do i=1,Ntot
       do irank=1,MpiSize-1
          send_offset(irank,i) = send_counts(irank-1,i) + send_offset(irank-1,i)
       enddo
    enddo
    !
    !master node is treated first.
    recv_offset(0,1) = 0
    do i=2,Ntot
       recv_offset(0,i) = sum(recv_counts(0,:i-1))
    enddo
    do i=1,Ntot
       do irank=1,MpiSize-1
          recv_offset(irank,i) = recv_offset(irank-1,i) + &
                                 sum(recv_counts(irank-1,:))
       enddo
    enddo
    !
    !Perform  the MPI_AllToAllV communication: 
    !for each column j held by the CPU, the CPU
    !m sends the element L of the column to the
    !CPU L which places it at position m, 
    !see Fig.7.
    do j=1,Ntot
       call MPI_AllToAllV(&
            A(:,j),&
            send_counts(:,j),&
            send_offset(:,j),&
            MPI_DOUBLE_PRECISION,&
            B(:,:),&
            recv_counts(:,j),&
            recv_offset(:,j),&
            MPI_DOUBLE_PRECISION,*
            MpiComm,ierr)
    enddo
    ! 
    !The transposed array is in row-major order. Local in-place
    !transposition is required to recover correct ordering of elements
    call local_transpose(B,ncol,qrow)
  end subroutine vector_transpose_MPI

  subroutine local_transpose(mat,nrow,ncol)
    integer                      :: nrow,ncol
    real(8),dimension(Nrow,Ncol) :: mat
    mat = transpose(reshape(mat,[Ncol,Nrow]))
    end subroutine local_transpose
\end{lstlisting}
The initial part of the procedure deals with the generic determination
of the input arrays required by the \texttt{MPI\_AllToAllV} procedure.   
In the first section (lines 17-29) the number of elements of the source buffer to be sent by each process
is determined. This information is stored  in the array
\texttt{send\_counts}. 
By construction, see \figu{figAll2All}, the corresponding amount of data received in the receive buffer,
i.e. \texttt{recv\_counts}, is obtained by transposing the
\texttt{send\_counts} array itself. This is performed using \texttt{MPI\_AllToAll} (lines
33-36). 
Next, the positions in memory of the send and receive elements have to be
determined. These correspond to the relative displacements of the
source and receive data  with respect to their starting memory addresses,
i.e. \texttt{send\_offset} and \texttt{recv\_offset},  which can be evaluated
by summing over the number of elements to be sent or received by each processor, see lines
42-57.

The matrix transposition is performed proceeding column-by-column, see
lines 65-76, repeatedly calling the  \texttt{MPI\_AllToAllV}
procedure~\cite{MPI2}.  
Note that some process may run out of columns to send because of the
incommensurability of the matrix size with the number of processes
(the parameter \texttt{Ntot} can be different from \texttt{Qcol}).
In this case MPI2 allows to send empty arrays.
The final array is however arranged in row-major mode, thus a local
transposition on each process is required to reorder the data in column-major mode.

Finally, we present the implementation of the parallel MVP procedure, as
described in \secu{sSecMPIall2all}: 
\begin{lstlisting}
  ! MPI_MatVec(Q,v,Hv) - $\ket{Hv} = H_\SS\ket{v}$
  subroutine spMatVec_mpi_main(Q,v,Hv)
    integer                          :: Q
    real(8),dimension(Q),intent(in)  :: v
    real(8),dimension(Q),intent(out) :: Hv
    ...
    !$H_d\ket{v}$ part
    Hv=0d0
    do i=1,Q
       do j=1,spH0d%row(i)%Size
          Hv(i) = Hv(i) + spH0d%row(i)%vals(j)*v(i)
       end do
    end do
    !
    !$\up$-part $H_\up\ket{v}$ - contiguous in memory.
    do idw=1,MpiQdw
       do iup=1,DimUp
          i = iup + (idw-1)*DimUp
          hxv_up: do jj=1,spH0ups(1)%row(iup)%Size
             jup = spH0ups(1)%row(iup)%cols(jj)
             jdw = idw
             val = spH0ups(1)%row(iup)%vals(jj)
             j   = jup + (idw-1)*DimUp
             Hv(i) = Hv(i) + val*v(j)
          end do hxv_up
       enddo
    end do
    !
    !$\dw$-part $H_\dw\ket{v}$ - non-contiguous in memory
    mpiQup=DimUp/MpiSize ; if(MpiRank<mod(DimUp,MpiSize))MpiQup=MpiQup+1
    allocate(vt(mpiQup*DimDw)) ;vt=0d0
    allocate(Hvt(mpiQup*DimDw));Hvt=0d0
    !
    !MPI vector transposition $\ket{v}\rightarrow \ket{v}^T=\ket{vt}$
    call vector_transpose_MPI(DimUp,MpiQdw,v,DimDw,MpiQup,vt)
    !
    !$\ket{Hvt} = H_\dw\ket{vt}$
    Hvt=0d0    
    do idw=1,MpiQup             !<= Transposed order:  column-wise DW <--> UP  
       do iup=1,DimDw           !<= Transposed order:  column-wise DW <--> UP
          i = iup + (idw-1)*DimDw
          hxv_dw: do jj=1,spH0dws(1)%row(iup)%Size
             jup = spH0dws(1)%row(iup)%cols(jj)
             jdw = idw             
             j   = jup + (jdw-1)*DimDw
             val = spH0dws(1)%row(iup)%vals(jj)
             Hvt(i) = Hvt(i) + val*vt(j)
          end do hxv_dw
       enddo
    end do
    deallocate(vt) ; allocate(vt(DimUp*mpiQdw)) ; vt=0d0
    !
    !Transpose back: $\ket{Hvt}^T=\ket{Hv}$
    call vector_transpose_MPI(DimDw,mpiQup,Hvt,DimUp,mpiQdw,vt)
    Hv = Hv + Vt
    deallocate(vt)
    !
    !Optional Non-Local terms: $H_{nd}\ket{v}$
    if(jhflag)then
       N = 0 ; call AllReduce_MPI(MpiComm,Q,N)
       allocate(vt(N)) ; vt = 0d0
       !Reconstructs the vector gathering all CPUs shares, see SETUP
       call allgather_vector_MPI(MpiComm,v,vt)
       !Each process get its own share of the product
       do i=1,Q
          matmul: do j=1,spH0nd%row(i)%Size
             Hv(i) = Hv(i) + spH0nd%row(i)%vals(j)*Vt(spH0nd%row(i)%cols(j))
          enddo matmul
       enddo
       deallocate(Vt)
    endif
    end
\end{lstlisting}

\subsection{Dynamical correlation functions }\label{CodeGF}
  The knowledge of the Hamiltonian spectrum, or just of its
lowest part, provides the means to evaluate generic dynamical correlation
functions of the form: $C_\OO = \ibra \TT_\pm[ \OO(t)
\OO^+ ]\iket$, where  $\OO(t)=e^{iHt}\OO e^{-iHt}$ is an operator in the Heisenberg
representation, $\TT_\pm$ is the time-ordering operator for fermions ($+$) or
bosons ($-$) and $\ibra \OO \iket=\frac{1}{Z}\Tr{ \left[ e^{-\beta
      H}\OO \right]}$,  $Z=\sum_ne^{-\beta E_n}$, is the thermodynamic
average.
Performing a Fourier transformation and using spectral theorem, the
expression for $C_\OO$ can be recasted in the more convenient form: 
\begin{equation}\label{KLgf}
  \begin{split}
    C_\OO(z) &=  \ibra \OO \frac{1}{z-H} \OO^+\iket
    \mp \ibra \OO^+ \frac{1}{z+H} \OO\iket \cr
    & = \frac{1}{Z}\sum_n e^{-\beta E_n}
    \bra{\psi_n}\OO \frac{1}{z-H}\OO^+ \ket{\psi_n}
    \mp
    \bra{\psi_n}\OO^+\frac{1}{z+H} \OO\ket{\psi_n}\cr
    & = \frac{1}{Z}\sum_n e^{-\beta E_n}\sum_m
    \frac{ \bra{\psi_n}\OO \ket{\psi_m} \bra{\psi_m}\OO^+ \ket{\psi_n} }  {z-(E_m-E_n)}
    \mp
    \frac{ \bra{\psi_n}\OO^+ \ket{\psi_m} \bra{\psi_m}\OO \ket{\psi_n} }  {z+(E_m-E_n)}\cr    
  \end{split}
\end{equation}
where $z\in\CCC$, 
$\ket{\psi_n}$ are the eigenstates of the
Hamiltonian $H$ and $E_0$ is the groundstate energy.
A key simplification in the evaluation of such dynamical correlation
function is obtained using the Lanczos procedure. Introducing a suitable Krylov basis 
the Hamiltonian matrix takes a tri-diagonal form, which allows to
evaluate an arbitrary number of terms in the sum over the 
internal excited states $\ket{\psi_m}$. 
Let's consider  the normalized initial state
$\ket{\phi_n}=\OO^+\ket{\psi_n}/\sqrt{\bra{\psi_n}\OO \OO^+\ket{\psi_n}}$ with
$\ket{\psi_n}\in\SS$ and the Krylov basis
$\mathcal{K}_{N} (\ket{\phi_n})=\{\ket{\phi_n}, H\ket{\phi_n}, \dots,
H^N\ket{\phi_n}\}\equiv \{\ket{v^n_0}, \ket{v^n_1},\dots, \ket{v^n_N}
\}$ with $1 \ll N \ll \DD_\SS$.
Any eigenstate $\ket{\psi_n}$ is decomposed along
the $\mathcal{K}_{N} (\ket{\phi_n})$ basis as: $\ket{\psi_n} =
\sum_i  \ibra v^n_i|\psi_n \iket  \ket{ v^n_i} =
\sum_i a^n_i \ket{v^n_i}$. Using this decomposition, we can evaluate
the operator components $\bra{\psi_m}\OO^+ \ket{\psi_n} = \sqrt{\bra{\psi_n}\OO
  \OO^+\ket{\psi_n}} {a^n_m}^*$ and thus:
\begin{equation}
  \bra{\psi_n}\OO \frac{1}{z-H}\OO^+ \ket{\psi_n} \simeq
  \bra{\psi_n}\OO \OO^+\ket{\psi_n}
  \sum_{m=1}^{N} \frac{ |a^n_m|^2}{ z - (E_m-E_n)}
\end{equation}

Repeating the same decomposition for each term in Eq.~\ref{KLgf}, we
obtain the following approximation to $G_\OO$:
\begin{equation}
  \begin{split}
    C_\OO(z)  &= \frac{1}{Z}\sum_n e^{-\beta E_n} C^n_\OO(z)\cr
    C^n_\OO(z)&\simeq \sum_{m=1}^{N} \frac{\bra{\psi_n}\OO \OO^+\ket{\psi_n} |a^n_m|^2}{ z - (E_m-E_n)} \mp \frac{\bra{\psi_n}\OO^+ \OO\ket{\psi_n} |a^n_m|^2}{ z + (E_m-E_n)}
  \end{split}
\end{equation}
  
In many concrete situations one has to deal with non-diagonal dynamical
correlation functions, e.g.: $C_{\AA\BB}(z) = \ibra A^+
\tfrac{1}{z-(H-E_0)} B\iket$, which can not be
treated by the procedure outlined above. A simple solution is to
consider auxiliary operators of the form $\OO=\AA +\BB$ and
$\PP=\AA-i\BB$. Using simple algebra it is then straightforward to
obtain the desired function $C_{\AA \BB}$ from the evaluation of
$C_\OO$ and $C_\PP$:
$$
C_{\AA\BB} = \frac{1}{2}\left[C_\OO + C_\PP - (1+i)C_\AA -(1+i)C_\BB\right]
$$

The evaluation of the impurity Green's functions $G_{\a\b}=\ibra \TT_\pm[ c(t)
c^+ ]\iket$  is executed in \texttt{GREENS\_NORMAL}.
Similarly, spin, charge, excitonic and pair susceptibilities functions are
evaluated in the modules \texttt{CHI\_SPIN}, \texttt{CHI\_DENS},
\texttt{CHI\_EXCT} and \texttt{CHI\_PAIR}.
All the dynamical correlation functions are evaluated along the
imaginary and real-frequency axis. The susceptibilities are also
evaluated in imaginary time. 
The most time consuming part of this step is the construction of the
Krylov basis $\mathcal{K}_N(\OO\ket{\psi_n})$ associated to each of the states
$\ket{\psi_n}$ of the spectrum determined in the diagonalization step. 
This ultimately requires to perform a number $N$ of MVP for any target
state $\ket{\psi_n}$.
This task is handled using parallel algorithm discussed above, greatly
reducing the computational cost of this operation. 
The order of the Krylov basis is controlled by the input variable \texttt{lanc\_gfniter}.

The evaluation of a dynamical
correlation function introduces an intrinsic small memory bottleneck,
related to the non-local nature of the operator application
$\ket{\phi_n}=\OO\ket{\psi_n}$, which in our implementation is
performed by the master node and subsequently distributed to the other
processes.

\subsection{Observables}\label{CodeObs}
Local observables, such as occupation, double occupancies and local
energy, are evaluated in the module \texttt{OBSERVABLES} using
thermal average  $\ibra \OO\iket = \Tr{e^{-\beta H}\OO}/Z$. 
Having access to the lowest part of the spectrum we can evaluate the
first terms in the thermodynamic trace: $\ibra \OO\iket 
= \tfrac{1}{Z}\sum_n e^{-\beta E_n}\bra{\psi_n}\OO\ket{\psi_n}$, where $\ket{\psi_n}$
are the  eigenstates of the system.

\subsection{Main}\label{SubSecMain}
The principal interfaces to the \NAME\ library are contained in the
\texttt{MAIN} module, which contains two procedures.
The \texttt{?init\_solver} aims to initialize the ED solver, by
performing a series of compatibility checks, allocating the required
static memory and initializing the bath class either performing a
generic guess or reading the bath from a file (specified by the input parameter
\texttt{Hfile}). This procedure accepts as only input the user defined
bath, i.e. an array of real numbers whose dimension
should conform the output of \texttt{?get\_bath\_dimension}, see
Table~\ref{TableProcedures}.   

The ED solver is interfaced by the procedure \texttt{?solve},
which requires as input the user bath array and the local
non-interacting Hamiltonian term $H^{loc}$ array.
The local Hamiltonian is used to setup of the structure of the
impurity problem.
A call to \texttt{?solve} performs all the calculations necessary to
solve the quantum impurity problem for the input bath. All the
dynamical correlations and observables are stored in the memory and
accessible via suitable input/output procedures, see Table~\ref{TableProcedures}. 

In presence of inequivalent atoms or sites the problem is solved 
by independently diagonalizing the impurity problem of each atom or site.
The user should allocate a rank-2 array to store the baths where the first dimension corresponds to the
number $N_{ineq}$ of inequivalent atoms or sites.
In analogy, the local non-interacting Hamiltonian should correspond to
a rank-5 array in which the first dimension should be $N_{ineq}$.
Upon call to \texttt{?solve}, each of the $N_{ineq}$ quantum impurity problems
is solved. Using the optional input parameter \texttt{mpi\_lanc} the user can
select parallel execution either on the inequivalent sites
(\texttt{mpi\_lanc}=False and default) or on the diagonalization algorithm
(\texttt{mpi\_lanc}=True).

\section{Installation and usage} \label{SecInstallUse}
The \NAME\ library is released under open source GNU GPL
license. The source code is publicly available at the GitHub on-line
repository:  
\href{https://github.com/QcmPlab/EDIpack}{github.com/QcmPlab/EDIpack}.
For the installation few dependencies have to be satisfied. In
particular the code relies on the open source scientific library
{SciFortran}, available at \href{https://github.com/QcmPlab/SciFortran}{github.com/QcmPlab/SciFortran}: 
\begin{itemize}
\item Fortran compiler with support to major fortran 90/2003
  standard, e.g. GNU gfortran version $>$ 9.0.0 or Intel ifort $>$ 13.0.0. 
\item CMake $>=$ 3.0.0
\item MPI
\item SciFortran 
\end{itemize}
Additional dependencies  for the Python binding are:
\begin{itemize}
\item Python, version $>=$3.6.0
\item Pkgconfig
\item Numpy
\item Mpi4py
\end{itemize}

\subsection{Installation}
The source code can be downloaded or cloned from the official
GitHub website. The build system is based on cmake. The library can be loaded into the operative system either via the provided shell script or as an environment module.  
Should all the dependencies be satisfied and assuming to be in the
main library directory, the compilation procedure is:
\begin{lstlisting}[language=bash,frame=none,numbers=none,basicstyle=\ttfamily]
mkdir Build
cd Build
cmake ..  [-DPREFIX=<~/opt/edipack/PLAT/VERSION> 
           -DBUILD_TYPE=<RELEASE>/TESTING/DEBUG 
           -DUSE_MPI=<yes>/no ] 
make 
make install
\end{lstlisting}

The Python binding can be installed once the library has been loaded
in the system using either:
\begin{lstlisting}[language=bash,frame=none,numbers=none,basicstyle=\ttfamily]
export F90=mpif90
pip install .
\end{lstlisting}

We tested installation of the \NAME\ library in different platforms,
e.g. common Linux distribution, Unix/OSX systems and dedicated HPC.

\subsection{Basic usage}\label{SecUsage}
The goal of \NAME\ is to provide an efficient and scalable method to
solve quantum impurity problems. As such, use of the library requires
the user to write a suitable driver program.
Two commented examples, using the Fortran and Python
interfaces respectively, are included in the  \texttt{test} directory of the
library.  These test programs aim to solve the DMFT equations for the single band
Hubbard model on the Bethe lattice. The provided codes can be used as
templates for other specific cases (see \href{https://github.com/QcmPlab/Driver-Database}{github.com/QcmPlab/DriverDatabase}).

\subsubsection{Fortran API}
The \NAME\ library is entirely written in Fortran. The internal structure of
the library is opaque to the user, so that only
a number of variables and procedures are available while the remaining
parts are not directly accessible by the user.
The interface is implemented by the  outmost module \texttt{EDIPACK},
which gives access to all the necessary procedures and control variables, see Table~\ref{TableParameters}
(first column) and Table~\ref{TableProcedures}.
The Fortran binding can be loaded in a program using the statement:
\begin{lstlisting}[language=fortran,frame=none,numbers=none,basicstyle=\ttfamily]
program TEST
  !Import EDIpack Fortran module
  USE EDIPACK
  ...
  implicit None
  ...
end program
\end{lstlisting}

Because Fortran does not allow namespace resolution, all the Fortran
procedures are preceded by the string \texttt{ed\_},
see Table~\ref{TableProcedures}.

  \subsubsection{Python API}\label{SecPythonApi}
Use of the \NAME\ library within Python is possible through suitable
API. The Python binding is built using a specific interface module
\texttt{edi2py} generated by \texttt{numpy.f2py} from a suitable
Fortran layer, i.e. a collection of Fortran interfaces linking to the
Fortran API. Finally, the Python module \texttt{edipy} is generated
operating on the  \texttt{edi2py} interface and exploiting
``duck typing'' to ensure full compatibility between Fortran and
Python API.
The Python binding can be loaded as:
\begin{lstlisting}[language=python,frame=none,numbers=none,basicstyle=\ttfamily]
#Import edipy, the EDIpack Python API
from edipy import *
import edipy as ed
\end{lstlisting}
The Python procedures are preceded by the namespace of the module as
\texttt{edipy.} and can be conveniently renamed as in this example or
in the available test code.

\subsubsection{MPI Parallel interface}
The \NAME\ library relies on distributed memory MPI
implementation~\cite{MPI2}.
The MPI specific code is included in the library
by means of pre-processing instructions identified by the macro
\texttt{\_MPI}. MPI support is activated by default in the installation.
The library supports  either parallel (MPI initialized) or serial (MPI not initialized) programs,
see \secu{SecUsage}. 

In order to keep the interface as simple as possible, the MPI support
is completely implicit. 
The parallel execution of any \NAME\ procedure is activated once the
MPI environment has been initialized in the calling program.
To this end, a system query \texttt{MPI\_Initialized} is used to
determine whether or not the user activated MPI. If true, MPI
instructions are executed and internal MPI setup is performed by
calling \texttt{ed\_set\_MpiComm}, contained in the module \texttt{VARS\_GLOBAL}.
Thus, the user is only required to  initialize (\texttt{MPI\_Init}) and finalize
(\texttt{MPI\_Finalize}) the MPI environment in the calling program in
order to perform parallel calculations. 
Note that in the Python module \texttt{mpi4py} these steps are
automatic. 
\subsubsection{Workflow}\label{SecWorkflow}
Although the calling program can differ substantially case by case, a
general workflow for the use of \NAME\ can be identified. Here we 
briefly discuss such workflow for the case of single-impurity
problem.
The case of inequivalent atoms or sites requires to
change the dimension of allocated array as discussed in
Table~\ref{TableProcedures}.   

The first part of the program requires to load \NAME\ and, if required, MPI should be initialized.
Next, the input file has to be read with a call to the procedure
\texttt{?read\_input}.
The only input argument of this procedure is the name of the input file.
Should the file not exist in the execution directory,  
a suitable input file with prefix \texttt{used.} is created, using
default values for all the input parameters along with a short
comment. After renaming, this file can be used as a starting point to setup the
calculation.
The \NAME\ control variables can additionally be controlled from
command line with the simple syntax \texttt{Variable=Value} appended after the program call. 
Moreover, a \textit{used} version of the input file with prefix
\texttt{used.}, containing the variables actually used during the run,
is written after each call to the main solver.
An overview of the input variables of the \NAME\ library is reported
in Table \ref{TableParameters}.
The introduction of many input parameters enables a higher level of
control over the library. Yet, their number can be frightening for the
inexperienced user. To this end, all the variables are initialized to
a default value which is suitable to perform calculation at zero
temperature for a single band model.
In the first part of Table \ref{TableParameters} we report the
most relevant input variables, controlling
important aspects of the impurity
problem including external parameters such as temperature, interaction
strength, chemical potential, etc.
In the second part table we list the remaining and more specific input
variables, which can be used to reach a finer tuning of the library
performances or to adapt the functioning to specific cases. 

After this step the user is required to setup some properties specific
to the problem.
For the DMFT application this usually reduces to generating the
necessary density of states $D(e)$ or the tight-binding Hamiltonian
$H(k)$.
The local part of the non-interacting Hamiltonian $H^{loc}$ should be allocated
and defined. The \NAME\ library requires this input to be a rank-4
dble array of shape
\texttt{[Nspin,Nspin,Norb,Norb]}. 
In addition, local dynamical function in the Matsubara axis, i.e. the
local Green's function $G(i\omega_n)$, or the impurity self-energy
$\Sigma(i\omega_n)$,  should be allocated with the correct dimensions.
Note that  \NAME\ requires these arrays to have shape \texttt{[Nspin,Nspin,Norb,Norb,Lmats]}, where
\texttt{Nspin}, \texttt{Norb} and \texttt{Lmats} are input variables.
Real-axis analogs of such functions can also be allocated and used,
although these do not necessarily participate to the self-consistency
procedure.

Terminated this part the user should initialize the solver.
This task requires two steps. The first is to query the \NAME\ library
to know the dimension of the rank-1  dble array storing the bath
parameters and used in the reverse communication implementation.
This information is retrieved with a call to 
\texttt{?get\_bath\_dimension}, which returns such dimension as an integer. 
The second step, is to call \texttt{?init\_solver} to initialize the
ED solver with the user bath array as the only input.   
On return, the input array contains the bath parameters.
These are either read from the file
\texttt{Hfile}.\texttt{restart}, if it exists in the executing
directory, or guessed from a default flat density of states controlled by the input
variables \texttt{ed\_hw\_bath} and \texttt{ed\_offset\_bath}, see
Table~\ref{TableParameters}.  

At this stage the user is ready to call the ED solver
\texttt{?solve} whose only inputs
are the bath and the local Hamiltonian arrays.
Within the DMFT framework this step is usually performed inside a
self-consistency cycle, which in its simplest form  just corresponds
to an iterative one.  
At each loop the user calls the \NAME\ ED solver, which solves the effective
quantum impurity problem determined by the user provided bath. This is
the most time-consuming part of the cycle. The results of the
calculations are stored in the memory up to a subsequent call of the
solver. Local dynamical impurity functions and many observables can be
retrieved from the memory using suitable procedures, see
Table~\ref{TableProcedures}.
In particular the impurity matsubara self-energy, used within the self-consistency, is retrieved with a call to
\texttt{?get\_sigma\_matsubara}.

Next, the user is required to perform the self-consistency condition,
which is highly dependent on the specific problem. This part of the
calling program is completely free and independent of \NAME.  

Finally, and in order to close the DMFT loop the user should update
the bath according to the outcome of the self-consistency equation.
This step can be performed using the conjugate gradient minimization provided with
\NAME\, i.e. calling \texttt{?chi2\_fitgf}.
Yet, other methods can in principle be adopted.
The updated bath parameters will be used as input of the ED solver,
until convergence is achieved.

\subsubsection{Output}
All the output of the \NAME\ library is written in plain text and
distributed among different files with the extension \texttt{.ed}.
The output can be divided in three main groups.

The first group contains all the dynamical impurity functions,
including the Green's functions, the self-energies and the
susceptibilities. The input variables \texttt{ed\_print\_Sigma},
\texttt{ed\_print\_G}, \texttt{ed\_print\_G0} can be used to allow or
suppress write out of, respectively, the self-energy, the impurity
and the non-interacting  Green's functions.
The quantities in this group are written to files of the form
\texttt{impFUNC\_l$\a\b$\_s$\sigma$\_x.ed} where \texttt{FUNC=G0, G,
  Sigma} indicates the non-interacting Green's function, the impurity
one and the self-energy, $\a$, $\b$ are
orbital indices, $\sigma=1,2$ corresponds to the spin index and
\texttt{x=iw, realw}, respectively for the Matsubara and real-axis frequency.
Each file contains three columns, in order: the frequency, the imaginary
part and the real part.

The second group of outputs contains the values of several static
quantities.
In order to ease the data analysis with non-blocking shell
commands such quantities are stored in multi-column files.
In order to avoid problems with data processing the columns headers
are reported in separate files using the same name of the data file
but with the suffix \texttt{\_info.ed}.
The most important observables and other significant quantities are
contained in the files   \texttt{observables\_all.ed} (which subsequently stores variables from
all calls to the solver) and \texttt{observables\_last.ed}
(which contains the values from the last call). 
The contributions to the local energy $\ibra H\iket$ are written in
columns in the file \texttt{energy\_last.ed}. 
The excitonic order parameters $P^m = \ibra c^+_{\a
  r}\sigma^m_{rs} c_{\b s}\iket$, with  $\sigma^{m=0,z}$ Pauli
matrices, are written to the file \texttt{exciton\_last.ed}.

The third group contains all the remaining files. These
include the bath used in the actual calculation, written to
\texttt{Hfile.used}, the restart files \texttt{Hfile.restart} and
\texttt{SectorFile.restart}, respectively containing the bath
resulting from the fit and the list of sectors QNs contributing to the
evaluated spectrum. The \texttt{.restart} files can be used to restart
a given calculation.   
The list of the eigenvalues determined during the diagonalization
procedure is written to \texttt{eigenvalues\_list.ed}. Similarly, a summary of
the states list is written to the file \texttt{state\_list.ed}.  
Some of the
parameters used in the calculations, conventionally used for data
analysis, are reported column-wise in the file
\texttt{parameters\_last.ed}.
The result of the fit is written to files of the form
\texttt{chi2fit\_results\_orb$\a$\_s$\sigma$.ed}, with $\a$ orbital index
and $\sigma=1,2$ the spin index. Similarly, the input function $X(i\omega_n)$ and
the fitted function $X^{And}(i\omega_n;\{V,h\})$ resulting from the
fit are written to files of the form
\texttt{fit\_X\_orb$\a$\_s$\sigma$.ed}, with $\a$ and $\sigma$  as above
and \texttt{X=Delta, Weiss} according to the value of
\texttt{cg\_scheme}. 

Finally, the \NAME and the SciFortran GIT SHA1 identifiers  used in the current calculation are
reported, respectively, in the files \texttt{EDIPACK\_version.inc} and
\texttt{scifor\_version.inc}.

\section*{Acknowledgements}
We acknowledge fruitful discussions with G.~Sangiovanni, A.~Sartori, H.~Choi, 
A.~Valli, S.~Adler and M.~Chatzieleftheriou. 
A.A. and M.C. acknowledge support from H2020 Framework Programme, 
under ERC Advanced Grant No. 692670 FIRSTORM and  financial support
from  MIUR PRIN 2015 (Prot. 2015C5SEJJ001) and SISSA/CNR project
``Superconductivity, Ferroelectricity and Magnetism in bad metals''
(Prot. 232/2015).
L.C. acknowledges financial support by the Deutsche
Forschungsgemeinschaft (DFG, German Research Foundation) under
Germany's Excellence Strategy through W\"urzburg‐Dresden Cluster of
Excellence on Complexity and Topology in Quantum Matter ‐ ct.qmat (EXC
2147, project‐id 390858490).
G.M. acknowledges support of the FNS/SNF through an Ambizione grant
and partly supported from the  European Research Council
(ERC-319286-QMAC).
L.d.M is supported by the European Commission through the ERC-CoG2016,
StrongCoPhy4Energy, GA No 724177.

\section*{Appendix A: Input Variables}\label{AppendixA}
Here is reported a list of the global input variables of the
library. The input parameters are arranged into the following table,
as discussed in \secu{SecWorkflow}.
Note that only a subset of the input parameters are transparent to the
user, i.e. can be directly accessed from the calling program.
These variables  are indicated with a symbol in the first column of
the table.  In the columns from $2^{\rm nd}$ to $4^{\rm th}$ we indicate,
respectively, the variable name, the type and the default value.
Finally, In the last column we provide a short
description of the variable purpose.


\begin{tabularx}{\linewidth}{ M{0.01\textwidth} T{0.19\textwidth}
    T{0.04\textwidth} D{0.11\textwidth} M{0.52\textwidth} }
  \caption{List of the \NAME\ input variables.}\\
  \hline 
  \hline
  {\it \footnotesize \phantom{T}} & {\it Variable} & {\it Type} & {\it Default} & {\it
                                                         \footnotesize
                                                         Description} \\
  \hline
  \hline
  \endhead
  \hline
  \multicolumn{5}{c}{\bf Most relevant control variables}\\
  \hline
  $\circ$ & Norb & Int & 1 & Number of impurity orbitals.\\
  \hline
  $\circ$ & Nspin& Int & 1 & Number of spin channels treated
                             independently. \texttt{Nspin}=1 enforces
                             non-magnetic solution.\\
  \hline
   & Nbath& Int & 6  & Number of bath sites. See \texttt{bath\_type}
                       for information about resulting total number of
                       electronic levels $N_s$.\\
  \hline
  $\circ$ & bath\_type& Char & ``normal''    & Set the bath type:
                                     \texttt{normal}:
                                     $N_s=\mathtt{Norb}(1+\mathtt{Nbath})$,
                                     \texttt{hybrid}:
                                     $N_s=\mathtt{Norb}+\mathtt{Nbath}$,
                                     and \texttt{replica}:
                                     $N_s=\mathtt{Norb}(1+\mathtt{Nbath})$,
                                     See \secu{SecBath}.\\
  \hline
  $\circ$ & Uloc& Dble & 2.0   & Local intra-orbital interactions $U$,
                                     different values per orbitals can
                                     be defined.\\
  \hline
  $\circ$ & Ust& Dble & 0.0   & Local inter-orbital interaction $U'$,
                                with opposite spin orientations.
                                For the Kanamori interaction this 
                                must be set to
                                $\mathtt{Uloc}-2\mathtt{Jh}$, with \texttt{Jh}
                                the Hund's coupling. 
                                The value $U''$ for equal spin orientation
                                is automatically set by the difference 
                                $\mathtt{Ust}-\mathtt{Jh}$. \\
  \hline
  $\circ$ & Jh   & Dble & 0.0 &  Hund's coupling.\\
  \hline
  $\circ$ & Jx   & Dble & 0.0 & Spin-Exchange coupling constant.\\
  \hline
  $\circ$ & Jp   & Dble & 0.0 & Pair-Hopping coupling constant.\\
  \hline
    $\circ$ & beta   & Dble & 1000 & Inverse temperature. For $T=0$
                                   calculations this value sets a 
                                   discretization of the Matsubara
                                   frequencies.\\
  \hline
  $\circ$ & xmu   & Dble & 0.0 & Chemical potential. This parameter
                                 sets the average occupations of the
                                 impurity. 
                                 If \texttt{HFMODE=True} this value
                                 contains the Hartree shift, i.e. $\mathtt{xmu}=0.0$ sets the half-filling for a particle-hole
                                 symmetric case.\\
  \hline
     & HFmode  & Bool & .true. & If True the Hartree shift is included in the
                interaction, e.g. $U(n_\up-1/2)(n_\dw-1/2)$. In this
                       case zero chemical potential corresponds to
                       half-filling for a particle-hole symmetric system.\\  
  \hline
  $\circ$ & Lmats & Int & 4096 & Number of Matsubara frequencies.\\
  \hline
  $\circ$ & Lreal  & Int & 5000 & Number of real-axis frequencies\\
  \hline
    $\circ$ & eps   & Dble & 0.01 & Real-axis broadening.\\
  \hline
  $\circ$ & wini, wfin & Dble & -5, 5 & Real-axis frequency range.\\
  \hline
  $\circ$ & Nloop    & Int & 100 & Maximum number of allowed iterations in the DMFT solution.\\
  \hline
  $\circ$ & dmft\_error & Dble & 1e-5 & Convergence threshold for the
                                        DMFT iterative cycle.\\
  \hline
  $\circ$ & Nph    & Int & 0 & Maximum number of phonons.\\
  \hline
  $\circ$ & g\_ph   & Dble & 0.0& Electron-Phonon coupling.\\
  \hline
  $\circ$ & w0\_ph  & Dble & 0.0& Holstein phonon frequency.\\
  \hline
  $\circ$ & xmin, xmax & Dble & -3, 3& Phonon probability distribution spatial range.\\
  \hline
   & chispin\_flag & Bool & .false. & Flag to include the evaluation of Spin-Spin susceptibilities.\\
  \hline
   &  chidens\_flag &Bool & .false. & Flag to include the evaluation of Density-Density susceptibilities.\\
  \hline
   & chipair\_flag & Bool & .false. &  Flag to include the evaluation
                                      of $s$-wave Pair susceptibilities.\\
  \hline
   & chiexct\_flag & Bool & .false. & Flag to include the evaluation of Excitonic susceptibilities.\\
  \hline
   & ed\_sparse\_H    & Bool & .true. &  Flag to select storage type of the sparse
                         Hamiltonian. True: sparse matrix object,
                         False:on-the-fly matrix-vector product.\\
  \hline
   & ed\_total\_ud     & Bool & .true. & Flag to select quantum numbers type.
                                 True: conserved total electron occupation per
                                 spin. 
                                 False: conserved electrons
                                 occupations per orbital and spin.\\
  \hline 
   & ed\_print\_Sigma & Bool & .true. & Flag to print out the impurity Self-energies.\\
  \hline
   & ed\_print\_G        & Bool & .true. & Flag to print out the interacting impurity Green's functions.\\
  \hline
   & ed\_print\_G0      & Bool & .true. & Flag to print out the non-interacting
                              impurity Green's functions.\\
  \hline
   & ed\_twin             & Bool & .false. & Flag to reduce the number of visited
                                       sector using  sectors symmetry:
                                       $[\vec{N}_\up,\vec{N}_\dw]\leftrightarrow[\vec{N}_\dw,\vec{N}_\up]$.
                                       A warning is raised if
                                       \texttt{ed\_twin=.true.} and
                                       \texttt{Nspin=2} as this
                                       symmetry 
                                       conflicts with possible long-range
                                       magnetic order.\\ 
  \hline
   & ed\_verbose       & Int & 3 & Verbosity level [0-5]. \\
  \hline
   & lanc\_method     & Char & ``arpack'' & String to select Lanczos
                                   method. \texttt{ARPACK}: uses
                                   P-Arpack, \texttt{LANCZOS}: uses
                                   single vector Lanczos method
                                   with no re-orthogonalization \\
  \hline
   & lanc\_tolerance  & Dble & 1e-18 & Tolerance for the Lanczos algorithm. \\
  \hline
   & lanc\_niter         & Int & 512 & Largest number of Lanczos
                                       iterations used in the
                                       Diagonalization of each sector Hamiltonian.\\
  \hline
   & lanc\_ngfiter      & Int & 200 & Largest dimension of the Krylov basis in
                              the evaluation of the Green's function.\\
  \hline
  & lanc\_nstates\_sector  & Int & 2 & Maximum number of eigenvalues per sector
                                        required to P-Arpack. For $T>0$ this number
                                        indicates an initial guess and
                                       is adjusted during the
                                       annealing.  \\
  \hline
   & lanc\_nstates\_total    & Int & 1 & Maximum number of states in
                                         the spectrum. 
                                         $\mathtt{lanc\_nstates\_total}=1$
                                         indicates that only the
                                         groundstate should be
                                         evaluated, resulting in zero
                                         temperature calculation.
                                         For $T>0$ calculations a value 
                                         $\mathtt{lanc\_nstates\_total}>1$
                                         is required. This value is
                                         used as an initial guess and adjusted
                                         during finite temperature
                                         annealing. See
                                         \texttt{cutoff} and
                                         Sec.~\ref{CodeEigenspace}. \\
  \hline
     & ed\_finite\_temp & Bool & .false. &  Flag to select finite temperature
                                         method. If True 
                                         $\mathtt{lanc\_nstates\_total} > 1$ is required.\\  
  \hline  

   & lanc\_ncv\_factor & Int & 10 & Factor to determine block size
                                    \texttt{Ncv} used in P-Arpack 
                                    using the expression:
                                          $\mathtt{Ncv}=\mathtt{lanc\_ncv\_factor}*\mathtt{Neigen}+\mathtt{lanc\_ncv\_add}$.
                                          P-Arpack requires \texttt{Ncv}
                                          to be at least twice of the
                                          required Eigensolutions \texttt{Neigen}.
                                          At zero temperature
                                          \texttt{Neigen}=\texttt{lanc\_nstates\_sector}. 
                                          At finite temperature
                                          \texttt{Neigen} is
                                          automatically adjusted from
                                          initial value, possibly leading to
                                          a block size \texttt{Ncv}
                                          too large.
                                          Reducing the value of
                                          \texttt{lanc\_ncv\_factor}
                                          down to its minimum value 2
                                          can be required to perform P-Arpack calculations.\\
  \hline
   & lanc\_ncv\_add       & Int & 0 & Factor to determine block size
                                      \texttt{Ncv} used in P-Arpack 
                                      using  the expression:
                                      $\mathtt{Ncv}=\mathtt{lanc\_ncv\_factor}*\mathtt{Neigen}+\mathtt{lanc\_ncv\_add}$.
                                      P-Arpack requires \texttt{Ncv}
                                      to be at least twice of the
                                      required Eigensolutions
                                      \texttt{Neigen}. \\
  \hline   
  $\circ$ & cg\_Scheme           & Char & ``Weiss'' & Conjugate Gradient fit scheme.
                                                      \texttt{Delta}: fit $\Delta(i\omega_n)$, \texttt{Weiss}: fit $\GG_0(i\omega_n)$.\\
  \hline
   & cg\_method           & Int & 0 & Conjugate Gradient minimization method.
                                      $\mathtt{cg\_method}=0$: Fletcher-Reeves-Polak-Ribiere
                                      minimisation algorithm. 
                                      $\mathtt{cg\_method}=1$: use the f77 \texttt{minimize} procedure published in
                                      Ref.\onlinecite{Georges1996RMP}
                                      and largely used in the DMFT community.\\
  \hline
   & cg\_Lfit   & Int & 1000 & Number of Matsubara frequencies used in
                               the calculation of the cost function
                               $\chi$ with the Conjugate
                               Gradient minimization. $\mathtt{Lfit}\leq\mathtt{Lmats}$.  \\
  \hline
   & cg\_grad             & Int & 0 & Type of gradient evaluation. $\mathtt{cg\_grad}=0$: Analytic,
                                      $\mathtt{cg\_grad}=1$:
                                      Numeric. If
                                      $\mathtt{cg\_method}=1$ then use
                                      of 
                                      $\mathtt{cg\_grad}=1$ is
                                      enforced as this method does not
                                      support analytic gradient calculation.\\
  \hline
  \hline
  \multicolumn{5}{c}{\bf More specific control variables, less
  frequently changed}\\
  \hline
  \hline  
  $\circ$ & sb\_field & Dble & 0.0& Small symmetry-breaking field.\\
  \hline  
  $\circ$ & Nsuccess  & Int & 1 & Minimum number of repeated iterations below
                                  threshold required to reach convergence.  \\
  \hline
   & cutoff  & Dble & 1e-9 & Cutoff for the Boltzmann factor in the
                             spectral summation, i.e.
                             $e^{-\beta(E-E_0)}<\mathtt{cutoff}$. See Sec.~\ref{CodeEigenspace}.\\
  \hline  
   & gs\_threshold & Dble & 1e-9 & Energy threshold for the groundstate
                      degeneracy. Eigenstates of energy $E$ within
                                   $|E-E_0|<\mathtt{gs\_threshold}$
                                   are considered as degenerate groundstates.\\
  \hline
   &ed\_sectors & Bool & .false. & This flag is used to reduce
                                         the number of sectors
                                    investigated during the
                                    construction of the spectrum. Only
                                    the sectors with QNs listed in the file
                                    \texttt{SectorFile} and those with
                                    plus and minus
                                    \texttt{ed\_sectors\_shift} are considered.\\
  \hline
    & ed\_sectors\_shift & Int & 1 & A shift of the QNs for the list
                                     of sectors read from
                                     \texttt{SectorFile} is
                                     \texttt{ed\_sectors=.true.}.
                                     The list is enlarged
                                     including the QNs obtained
                                     by applying a shift
                                     $\pm
                                     1,\ldots,\pm\mathtt{ed\_sectors\_shift}$
                                     to each element of the list.\\
  \hline
   & ed\_hw\_bath & Dble & 2.0 & The half-bandwidth for the bath
                              initialization using discretized flat density of
                              states.\\
  \hline  
   & ed\_offset\_bath & Dble & 0.1 & Energy offset for the
                                   initialization of the diagonal
                                   terms in replica bath. A small
                                   value randomly chosen in the
                                   interval
                                   $\epsilon\in [-\mathtt{ed\_offset\_bath}:\mathtt{ed\_offset\_bath}]$
                                   is added to the diagonal terms of
                                   the bath to avoid possible spectral
                                     degeneracies.\\
  \hline
  $\circ$ & Lpos & Int & 100 & Number of points used to obtain the phonon probability
                     distribution function.\\ 
  \hline
  $\circ$ & nread & Dble & 0.0 & Target density for fixed density
                       calculations. Use the procedure
                       \texttt{ed\_search\_variable}   to adjust
                       chemical potential \texttt{xmu} so that local
                       occupation is
                       $|n-\mathtt{nread}|<\mathtt{nerr}$. \\
  \hline
   & nerr & Dble & 1e-4 & Error threshold for fixed-density
                       calculations.\\
  \hline 
   & ndelta& Dble & 0.1 & Initial step for the variation of the chemical
                       potential in fixed-density calculations.\\
  \hline  
   & ncoeff & Dble & 1d0 & A multiplier for the initial ndelta as read
                           from a file: $\mathtt{ndelta}\rightarrow\mathtt{ndelta}*\mathtt{ncoeff}$.\\
  \hline  
   & ed\_solve\_offdiag\_gf & Bool & .false. & Force the calculation of
                                         the off-diagonal Green's
                                         functions.
                                         True if \texttt{bath\_type}$\neq$\texttt{normal} \\ 
  \hline
   & ed\_all\_G           & Bool & .false. & Flag to evaluate all components of the
                                             impurity Green's functions,
                                             irrespective of the symmetries of the
                                             problem when using
                                             \texttt{bath\_type}=\texttt{replica}.
                                             If False only the Green's
                                             functions corresponding
                                             to non-vanishing
                                             components of the bath
                                             Hamiltonian
                                             $h^\nu_{\a\b\sigma}$ are
                                             evaluated.  \\
  \hline
   & lanc\_nstates\_step    & Int & 2 & Number of states to be added
                                        to
                                        \texttt{lanc\_nstates\_sector}
                                        for each sector during finite
                                        temperature annealing of the
                                        state list. See
                                        Sec.~\secu{CodeEigenspace}. \\  
  \hline
   & lanc\_dim\_threshold   & Int & 1024 & Dimension threshold below
                                           which Lapack diagonalization method is
                                           used to determine
                                           eigensolutions. \\
  \hline
   & cg\_Niter             & Int & 500 & Maximum number of iteration
                                         for the Conjugate Gradient minimization.\\
  \hline
   & cg\_Ftol             & Dble & 1e-6 & Tolerance for the Conjugate Gradient minimization.\\
  \hline
   & cg\_stop             & Int & 0 & Stop condition for the Conjugate Gradient
                                      minimization of the cost function $\chi(x)$.
                                      0: $C_1+C_2$, 1: $C_1$, 2: $C_2$ with
                                      conditions 
                                      $C_1=|\chi^{n-1} - \chi^n|<\mathtt{cg\_Ftol}(1+\chi^n)$,
                                      $C_2=||x^{n-1}
                                      -x^n||<\mathtt{cg\_Ftol}(1+||x^n||)$. See
                                      Sec.~\secu{SecBathFit}.\\
  \hline
   & cg\_Weight         & Int & 1 & A weight factor used in the cost
                                    function $\chi$ of the Conjugate
                                    Gradient minimization. Increasing this
                                    parameter favors the fit of the low
                                    frequency part of $\chi$.
                                    1: $\mathtt{cg\_weight}=1$,
                                    2: $\mathtt{cg\_weight}=1/\mathtt{Lfit}$,
                                    3:
                                    $\mathtt{cg\_weight}=1/\omega_n$. See
                                    Sec.~\secu{SecBathFit}.\\
  \hline
   & cg\_pow             & Int & 2 & The exponent of the cost function
                                     $\chi$ as 
                                     $|X -
                                     X^{And}|^{\mathtt{cg\_pow}}$ used
                                     in Conjugate Gradient
                                     minimization. See Sec.~\ref{SecBathFit}.\\
  \hline
  $\circ$ & Hfile & Char & ``hamiltonian'' & File name to read/write bath parameters\\
  \hline
  & SectorFile & Char & ``sectors'' & File name to read/write
                                             sectors contributing to
                                             the state list. On output
                                             it contains the QNs of
                                             the list of sectors
                                             contributing to the
                                             evaluated spectrum. On
                                             input such list is used
                                             to reduce the sectors
                                             analysis. See
                                      \texttt{ed\_sectors} and \texttt{ed\_sectors\_shift}. \\
  \hline
  $\circ$ & LOGfile  & Int & 6 & Fortran unit for the log
                                 output. Value must be in  the range
                                 $[6:999]$. 6 is the standard output. \\
  \hline
  \label{TableParameters}
\end{tabularx}

\section*{Appendix B: Main procedures}\label{AppendixB}
In this appendix we report a list of all the procedures made
accessible by either Fortan or Python API of the \NAME\ library. As
discussed in Sec.~\ref{SecUsage} all such procedures feature implicit
MPI support, i.e. they are executed in parallel if the MPI framework
has been initialized in the calling program.

In the first column we report the procedure names and the possible
input arguments. In order to unify the naming conventions of the
interfaces we use the prefix \texttt{?=ed\_,  edipack.} to indicate,
respectively, the Fortran and Python procedures.
In the second column we briefly describe the purpose of
the procedure and the nature of the arguments.
The inputs in square brackets are optional.

The procedures are grouped with respect to the library module
containing them, indicated in a single column at the beginning of each
group.

\begin{tabularx}{\linewidth}{ T{0.46\textwidth}  M{0.49\textwidth} }
  \caption{List of the \NAME\ procedures.}\\
  \hline
  \hline
  {\it PROCEDURE} {(\texttt{?=ed\_, edipack.})} & {\it \footnotesize INFO} \\
  \hline
  \hline
  \endhead
  \hline
  \hline
  \multicolumn{2}{c}{\bf INPUT\_VARS}\\
  \hline
  \hline
  ?read\_input(File) &
                       Reads global input variables from the user
                       indicated file \texttt{File}.
                       Argument: \texttt{File} is a character string of arbitrary length.
                       If the file does not exist in the calling directory
                       a default one with prefix 
                       ``\texttt{used.}'' is produced using default
                       values.
                       The value of
                       each variable can be updated also from command line
                       using the syntax \texttt{Variable=}$Value$.\\
  \hline
  \hline
  \multicolumn{2}{c}{\bf  MAIN}\\
  \hline
  \hline
  ?init\_solver(Bath) & Initializes the whole ED calculation, allocate global static 
                        memory, perform compatibility checks and setup global
                        variables.
                        Argument: \texttt{Bath}  is a rank-1 dble array 
                        of dimension \texttt{Nb=?get\_bath\_dimension}.
                        For problems with \texttt{Nsites} inequivalent atoms
                        \texttt{Bath} is a rank-2 dble array of 
                        dimensions [\texttt{Nsites},\texttt{Nb}].
                        On \texttt{Bath} contains the bath parameters, either guessed from
                        flat density of states  or read from the
                        file \texttt{Hfile.restart}. \\
  \hline
  ?solve(Bath,Hloc)  &   Solve the quantum impurity problem using ED
                       method.
                       Arguments: \texttt{Bath} is a rank-1 dble array
                        of dimension
                       \texttt{Nb=?get\_bath\_dimension}.
                       For problems with \texttt{Nsites} inequivalent atoms
                       \texttt{Bath} is a rank-2 dble array of 
                       dimensions [\texttt{Nsites},\texttt{Nb}].
                       \texttt{Hloc} is a rank-4 dble array of
                       dimensions [\texttt{Nspin,Nspin,Norb,Norb}] or
                       a rank-5 of dimensions
                       [\texttt{Nsites,Nspin,Nspin,Norb,Norb}]. Arguments
                       are not changed on output.\\

  \hline
  \hline
  \multicolumn{2}{c}{\bf  IO}\\
  \hline
  \hline
  ?get\_sigma\_matsubara(Func,[Nsites])  & Returns the Matsubara
                                              self-energy function in \texttt{Func}.
                                              Argument: \texttt{Func}
                                              is a rank-5 cmplx array
                                              of dimensions
                                              [\texttt{Nspin,Nspin,Norb,Norb,Lmats}].
                                              If \texttt{Nsites} is
                                              present \texttt{Func}
                                              is a rank-6 cmplx array
                                              of dimensions
                                              [\texttt{Nsites,Nspin,Nspin,Norb,Norb,Lmats}]
                                              and the procedure returns
                                              the function for each
                                              inequivalent atom.\\
  \hline
  ?get\_sigma\_realaxis(Func,[Nsites])  &  Returns the real-axis
                                              self-energy function in \texttt{Func}.
                                              Argument: \texttt{Func}
                                              is a rank-5 cmplx array
                                              of dimensions
                                              [\texttt{Nspin,Nspin,Norb,Norb,Lreal}].
                                              If \texttt{Nsites} is
                                              present \texttt{Func}
                                              is a rank-6 cmplx array
                                              of dimensions
                                              [\texttt{Nsites,Nspin,Nspin,Norb,Norb,Lreal}]
                                              and the procedure returns
                                              the function for each
                                              inequivalent atom.\\
  \hline
  ?get\_gimp\_matsubara(Func,[Nsites])  & Returns the Matsubara
                                             impurity Green's function in \texttt{Func}.
                                              Argument: \texttt{Func}
                                              is a rank-5 cmplx array
                                              of dimensions
                                              [\texttt{Nspin,Nspin,Norb,Norb,Lmats}].
                                              If \texttt{Nsites} is
                                              present \texttt{Func}
                                              is a rank-6 cmplx array
                                              of dimensions
                                              [\texttt{Nsites,Nspin,Nspin,Norb,Norb,Lmats}]
                                              and the procedure returns
                                              the function for each
                                              inequivalent atom.\\
  \hline
  ?get\_gimp\_realaxis(Func,[Nsites])  & Returns the real-axis
                                             impurity Green's function in \texttt{Func}.
                                              Argument: \texttt{Func}
                                              is a rank-5 cmplx array
                                              of dimensions
                                              [\texttt{Nspin,Nspin,Norb,Norb,Lreal}].
                                              If \texttt{Nsites} is
                                              present \texttt{Func}
                                              is a rank-6 cmplx array
                                              of dimensions
                                              [\texttt{Nsites,Nspin,Nspin,Norb,Norb,Lreal}]
                                              and the procedure returns
                                              the function for each
                                              inequivalent atom.\\
  \hline

  ?get\_dens(Var,[Nsites])  & Returns the impurity occupations in
                                 \texttt{Var}.
                                 Argument: \texttt{Var} is a rank-1 dble array of dimensions [\texttt{Norb}].
                                 If \texttt{Nsites} is present \texttt{var} is a rank-2 dble array of
                                 dimensions [\texttt{Nsites,Norb}] and the procedure returns
                                 the value for each inequivalent atom.\\
  \hline
  ?get\_mag(Var,[Nsites])  &  Returns the impurity magnetization in
                                \texttt{Var}.
                                Argument: \texttt{Var} is a rank-1 dble array of dimensions [\texttt{Norb}].
                                If \texttt{Nsites} is present \texttt{var} is a rank-2 dble array of
                                dimensions [\texttt{Nsites,Norb}] and the procedure returns
                                the value for each inequivalent atom.\\
  \hline
  ?get\_docc(Var,[Nsites])  & Returns the impurity double occupancy
                                 in \texttt{Var}.
                                 Argument: \texttt{Var} is a rank-1 dble array of dimensions [\texttt{Norb}].
                                 If \texttt{Nsites} is present \texttt{var} is a rank-2 dble array of
                                 dimensions [\texttt{Nsites,Norb}] and the procedure returns
                                 the value for each inequivalent atom.\\
  \hline
  ?get\_eimp(Var,[Nsites])  & Returns the impurity local energies components
                                 in \texttt{Var}.
                                 Argument: \texttt{Var} is a rank-1 dble array of dimensions [4].
                                 If \texttt{Nsites} is present \texttt{var} is a rank-2 dble array of
                                 dimensions [\texttt{Nsites,4}] and the procedure returns
                                 the value for each inequivalent atom.\\
  \hline
  ?get\_doubles(Var,[Nsites])  & Returns additional impurity double
                                 occupancies from multi-orbital
                                 terms in \texttt{Var}.
                                 Argument: \texttt{Var} is a rank-1 dble array of dimensions [4].
                                 If \texttt{Nsites} is present \texttt{var} is a rank-2 dble array of
                                 dimensions [\texttt{Nsites,4}] and the procedure returns
                                 the value for each inequivalent atom.\\

  \hline
  \hline
  \multicolumn{2}{c}{\bf  BATH}\\
  \hline
  \hline
  ?get\_bath\_dimension() & Returns the number \texttt{Nb} of bath parameters
                               required to describe the quantum
                               impurity problem, based on the value of
                               different input variables,
                               e.g. \texttt{Norb, Nbath, bath\_type}.
                               Output: \texttt{Nb} integer constant. \\
  \hline
  \makecell[l]{
  ?set\_Hreplica(Hloc) \\
  ?set\_Hreplica(Hvec,LambdaVec)} & Sets up the matrix basis and the
                                    initial guess for the variational
                                    bath parameters which determined
                                    the shape of the  bath for \texttt{bath\_type=replica}.
                                    The procedure
                                    accepts either the local
                                    non-interacting Hamiltonian
                                    \texttt{Hloc} or  the \texttt{Nsym}
                                    matrix basis components
                                    \texttt{Hvec} and the
                                    corresponding variational
                                    parameters \texttt{Lambdavec}. See
                                    Sec.~\ref{SecBath}.  
                                    Arguments: \texttt{Hloc} is a
                                    rank-4 dble array of dimensions
                                    [\texttt{Nspin,Nspin,Norb,Norb}]
                                    or a rank-2 dble array of dimensions
                                    [\texttt{Nspin*Norb,Nspin*Norb}].
                                    \texttt{Hvec} is a rank-5 dble
                                    array of dimensions
                                    [\texttt{Nspin,Nspin,Norb,Norb,Nsym}].
                                    \texttt{LambdaVec} is a rank-1
                                    dble array of dimensions
                                    \texttt{Nsym} or a rank-2 dble
                                    array of dimensions
                                    [\texttt{Nsites,Nsym}], for
                                    \texttt{Nsites} inequivalent
                                    atoms. \\
  \hline
  
  ?get\_bath\_component\_dimension(Type) & Returns the dimensions a 
                                           rank-3 dble array should
                                           have to store a specific
                                           component of the bath. 
                                           The output is a rank-1 int
                                           array \texttt{Ndim(1:3)} of dimension 3. 
                                           Argument: \texttt{Type} is
                                           a single
                                           character. Possible values
                                           are \texttt{Type=e, v, l},
                                           corresponding to energy,
                                           hybridization or lambda
                                           components. \\
  \hline
  ?get\_bath\_component(Array,Bath,Type) & Returns in
                                                \texttt{Array} the
                                                specified components
                                                \texttt{Type} of the
                                                bath \texttt{Bath}.
                                                Arguments:
                                                \texttt{Array} a
                                                rank-3 dble array of
                                                dimensions
                                                [\texttt{Ndim(1),Ndim(2),Ndim(3)}]
                                                as returned by a call
                                                to
                                                \texttt{?get\_bath\_component\_dimension(Type)}.
                                                \texttt{Bath} is a
                                                rank-1 dble array of dimension
                                                \texttt{Nb=?get\_bath\_dimension}.
                                                \texttt{Type=e, v, l}
                                                is  a single character.\\
  \hline
  ?set\_bath\_component(Array,Bath,Type) &
                                               Sets the specified component
                                               \texttt{Type} in the
                                               bath \texttt{Bath}
                                               to \texttt{Array}.
                                               Arguments:
                                               \texttt{Array} a
                                               rank-3 dble array of
                                               dimensions
                                               [\texttt{Ndim(1),Ndim(2),Ndim(3)}]
                                               as returned by a call
                                               to
                                               \texttt{?get\_bath\_component\_dimension(Type)}.
                                               \texttt{Bath} is a
                                                rank-1 dble array of dimension
                                               \texttt{Nb=?get\_bath\_dimension}.
                                               \texttt{Type=e, v, l}
                                               is  a single character.\\
  \hline
  ?copy\_bath\_component(BathIN,BathOUT,Type) & Copies the specified component
                                               \texttt{Type} from the
                                                   input bath
                                                   \texttt{BathIn} to
                                                   the output bath \texttt{BathOut}. 
                                                   Arguments:
                                                   \texttt{BathIn} and
                                                   \texttt{BathOut}
                                                   are rank-1 dble array of dimension
                                                   \texttt{Nb=?get\_bath\_dimension}.
                                                   \texttt{Type=e, v, l}
                                                   is  a single
                                                   character.\\
  
  \hline
  ?spin\_symmetrize\_bath(Bath[,save])& Enforces spin symmetry in the bath
                                        \texttt{Bath}.
                                        Arguments: \texttt{Bath} is a rank-1 dble array
                                        of dimension
                                        \texttt{Nb=?get\_bath\_dimension}.
                                        For problems with \texttt{Nsites} inequivalent atoms
                                        \texttt{Bath} is a rank-2 dble array of 
                                        dimensions [\texttt{Nsites},\texttt{Nb}].
                                        \texttt{save} is a logical
                                        value used to optionally write
                                        out the output bath.\\
  \hline
  ?orb\_symmetrize\_bath(Bath[,save])& Enforces symmetry among orbitals in the bath
                                   \texttt{Bath}.
                                   Arguments: \texttt{Bath} is a rank-1 dble array
                                        of dimension
                                        \texttt{Nb=?get\_bath\_dimension}.
                                        For problems with \texttt{Nsites} inequivalent atoms
                                        \texttt{Bath} is a rank-2 dble array of 
                                        dimensions [\texttt{Nsites},\texttt{Nb}].
                                        \texttt{save} is a logical
                                        value used to optionally write
                                        out the output bath.\\
  \hline
  ?orb\_equality\_bath(Bath,Indx[,save])&
                                             Equals each orbital 
                                             components in the bath
                                             \texttt{Bath} to be
                                             identical to those of the
                                             orbital \texttt{Indx}.
                                              Arguments: \texttt{Bath} is a rank-1 dble array
                                             of dimension
                                             \texttt{Nb=?get\_bath\_dimension}.
                                             For problems with \texttt{Nsites} inequivalent atoms
                                             \texttt{Bath} is a rank-2 dble array of 
                                             dimensions
                                             [\texttt{Nsites},\texttt{Nb}].
                                             \texttt{Indx} is a
                                             integer constant. 
                                             \texttt{save} is a logical
                                             value used to optionally write
                                             out the output bath.\\
  \hline
  ?ph\_symmetrize\_bath(Bath[,save])& Enforces particle-hole symmetry
                                      in the bath \texttt{Bath}.
                                       Arguments: \texttt{Bath} is a rank-1 dble array
                                        of dimension
                                        \texttt{Nb=?get\_bath\_dimension}.
                                        For problems with \texttt{Nsites} inequivalent atoms
                                        \texttt{Bath} is a rank-2 dble array of 
                                        dimensions [\texttt{Nsites},\texttt{Nb}].
                                        \texttt{save} is a logical
                                        value used to optionally write
                                        out the output bath.\\
  \hline
  ?ph\_trans\_bath(Bath[,save])& Perform a particle-hole transformation onto the
                          bath \texttt{Bath}.
                           Arguments: \texttt{Bath} is a rank-1 dble array
                                        of dimension
                                        \texttt{Nb=?get\_bath\_dimension}.
                                        For problems with \texttt{Nsites} inequivalent atoms
                                        \texttt{Bath} is a rank-2 dble array of 
                                        dimensions [\texttt{Nsites},\texttt{Nb}].
                                        \texttt{save} is a logical
                                        value used to optionally write
                                        out the output bath.\\
  \hline
  ?break\_symmetry\_bath(Bath,Field,Sign[,save]) &
                                                    Breaks spin
                                                    symmetry in the bath \texttt{Bath}
                                                   using a small field of
                                                   amplitude 
                                                   \texttt{Field} with sign
                                                   \texttt{Sign}. The
                                                   quantity
                                                   \texttt{Field*Sign}
                                                   is added or
                                                   subtracted,
                                                   respectively for spin
                                                   up and down, to the
                                                   local energies of each bath level.  
                                                   Arguments: \texttt{Bath} is a rank-1 dble array
                                                   of dimension
                                                   \texttt{Nb=?get\_bath\_dimension}.
                                                   For problems with \texttt{Nsites} inequivalent atoms
                                                   \texttt{Bath} is a rank-2 dble array of 
                                                   dimensions
                                                   [\texttt{Nsites},\texttt{Nb}].
                                                   \texttt{Field,
                                                   Sign} are dble
                                                   constant.                                                     
                                                   \texttt{save} is a logical
                                                   value used to optionally write
                                                   out the output bath.\\
  \hline

  \hline
  \hline
  \multicolumn{2}{c}{\bf  BATH\_FIT}\\
  \hline
  \hline
  ?chi2\_fitgf(Func,Bath,Hloc,ispin[,iorb])   &  Optimizes the bath
                                                \texttt{Bath}
                                                according to the user-provided function
                                                \texttt{Func}, using a  Conjugate
                                                Gradient minimization
                                                of the cost function
                                                $\chi$, see
                                                Sec.~\ref{SecBathFit}.
                                                Argument: \texttt{Func}
                                                is a rank-5 cmplx array
                                                of dimensions 
                                                [\texttt{Nspin,Nspin,Norb,Norb,:}].
                                                For problems with
                                                \texttt{Nsites}
                                                inequivalent atoms
                                                \texttt{Func} is a rank-6 cmplx array
                                                of dimensions
                                                [\texttt{Nsites,Nspin,Nspin,Norb,Norb,:}]. 
                                                \texttt{Bath} is a rank-1 dble array
                                                of dimension
                                                \texttt{Nb=?get\_bath\_dimension}.
                                                or  a rank-2 dble array of 
                                                dimensions
                                                [\texttt{Nsites},\texttt{Nb}].
                                                \texttt{Hloc} is a rank-4 dble array of
                                                dimensions [\texttt{Nspin,Nspin,Norb,Norb}] or
                                                a rank-5 of dimensions
                                                [\texttt{Nsites,Nspin,Nspin,Norb,Norb}].
                                                \texttt{ispin} is an
                                                int constant
                                                indicating the spin
                                                components to fit.
                                                The optional argument
                                                \texttt{iorb} indicate
                                                the orbital components
                                                to fit. If not passed
                                                all orbitals are
                                                fitted.  \\   
  \hline

   \hline
  \hline
  \multicolumn{2}{c}{\bf  AUX\_FUNX}\\
  \hline
  \hline
  ?set\_suffix(ilat,pads)   &  Sets an additional suffix for the
                              output files of the \NAME\ library.
                              Arguments: \texttt{ilat} is an integer
                              constant. \texttt{pads} is an integer
                              constant indicating the number of zero
                              padding of the integer \texttt{ilat} in
                              the suffix.\\
  \hline
  ?reset\_suffix()   &  Resets the suffix for the output
                                   files of \NAME\ to the default value \\
  \hline
  ?search\_variable(var,ntmp,bool)   &  Varies the variable
                                         \texttt{var} so that the
                                       conjugated variable
                                       \texttt{ntmp} is equal to the
                                       input value \texttt{nread} up
                                       to an error \texttt{nerr}.
                                       Arguments: \texttt{var, ntmp}
                                       are dble
                                       constants. \texttt{bool} is a
                                       logical constant indicating the success of the variation.\\
  \hline
  \label{TableProcedures}
\end{tabularx}

\bibliographystyle{elsarticle-num}
\bibliography{references}

\begin{thebibliography}{10}
\expandafter\ifx\csname url\endcsname\relax
  \def\url#1{\texttt{#1}}\fi
\expandafter\ifx\csname urlprefix\endcsname\relax\def\urlprefix{URL }\fi
\expandafter\ifx\csname href\endcsname\relax
  \def\href#1#2{#2} \def\path#1{#1}\fi

\bibitem{Lin1993CIP}
H.~Lin, J.~Gubernatis, \href{https://doi.org/10.1063/1.4823192}{Exact
  diagonalization methods for quantum systems}, Computers in Physics 7~(4)
  (1993) 400--407.
\newline\urlprefix\url{https://doi.org/10.1063/1.4823192}

\bibitem{Hewson1993}
A.~C. Hewson, {The Kondo Problem to Heavy Fermions}, Cambridge University
  Press, New York, N.Y., 1993.

\bibitem{Kotliar2004PT}
G.~Kotliar, D.~Vollhardt, \href{https://doi.org/10.1063/1.1712502}{Strongly
  correlated materials: Insights from dynamical mean-field theory}, Physics
  Today 57~(3) (2004) 53--59.
\newline\urlprefix\url{https://doi.org/10.1063/1.1712502}

\bibitem{Georges1996RMP}
A.~Georges, G.~Kotliar, W.~Krauth, M.~J. Rozenberg,
  \href{https://link.aps.org/doi/10.1103/RevModPhys.68.13}{Dynamical mean-field
  theory of strongly correlated fermion systems and the limit of infinite
  dimensions}, Rev. Mod. Phys. 68 (1996) 13--125.
\newblock \href {https://doi.org/10.1103/RevModPhys.68.13}
  {\path{doi:10.1103/RevModPhys.68.13}}.
\newline\urlprefix\url{https://link.aps.org/doi/10.1103/RevModPhys.68.13}

\bibitem{Georges2004ACP}
A.~Georges, \href{https://aip.scitation.org/doi/abs/10.1063/1.1800733}{Strongly
  correlated electron materials: Dynamical mean‐field theory and electronic
  structure}, AIP Conference Proceedings 715~(1) (2004) 3--74.
\newblock \href {https://doi.org/10.1063/1.1800733}
  {\path{doi:10.1063/1.1800733}}.
\newline\urlprefix\url{https://aip.scitation.org/doi/abs/10.1063/1.1800733}

\bibitem{Lechermann2006PRB}
F.~Lechermann, A.~Georges, A.~Poteryaev, S.~Biermann, M.~Posternak,
  A.~Yamasaki, O.~K. Andersen,
  \href{http://link.aps.org/abstract/PRB/v74/e125120}{{Dynamical mean-field
  theory using Wannier functions: A flexible route to electronic structure
  calculations of strongly correlated materials}}, Phys. Rev. B 74~(12) (2006)
  125120.
\newblock \href {https://doi.org/10.1103/PhysRevB.74.125120}
  {\path{doi:10.1103/PhysRevB.74.125120}}.
\newline\urlprefix\url{http://link.aps.org/abstract/PRB/v74/e125120}

\bibitem{Potthoff2003TEPJBCMACS}
M.~Potthoff,
  \href{https://doi.org/10.1140/epjb/e2003-00121-8}{Self-energy-functional
  approach to systems of correlated electrons}, The European Physical Journal B
  - Condensed Matter and Complex Systems 32~(4) (2003) 429--436.
\newblock \href {https://doi.org/10.1140/epjb/e2003-00121-8}
  {\path{doi:10.1140/epjb/e2003-00121-8}}.
\newline\urlprefix\url{https://doi.org/10.1140/epjb/e2003-00121-8}

\bibitem{Senechal2008}
D.~{Sénéchal}, The variational cluster approximation for hubbard models:
  Practical implementation, in: 2008 22nd International Symposium on High
  Performance Computing Systems and Applications, 2008, pp. 9--15.

\bibitem{Potthoff2011ACP}
M.~Potthoff, \href{https://aip.scitation.org/doi/abs/10.1063/1.3667325}{Static
  and dynamic variational principles for strongly correlated electron systems},
  AIP Conference Proceedings 1419~(1) (2011) 199--258.
\newblock \href {https://doi.org/10.1063/1.3667325}
  {\path{doi:10.1063/1.3667325}}.
\newline\urlprefix\url{https://aip.scitation.org/doi/abs/10.1063/1.3667325}

\bibitem{Nuss2011}
M.~Nuss, E.~Arrigoni, M.~Aichhorn, W.~von~der Linden, Variational cluster
  approach to the single impurity anderson model (2011).
\newblock \href {http://arxiv.org/abs/1110.4533} {\path{arXiv:1110.4533}}.

\bibitem{Bauer2011JOSMTAE}
B.~Bauer, L.~D. Carr, H.~G. Evertz, A.~Feiguin, J.~Freire, S.~Fuchs, L.~Gamper,
  J.~Gukelberger, E.~Gull, S.~Guertler, A.~Hehn, R.~Igarashi, S.~V. Isakov,
  D.~Koop, P.~N. Ma, P.~Mates, H.~Matsuo, O.~Parcollet, G.~Pawlowski, J.~D.
  Picon, L.~Pollet, E.~Santos, V.~W. Scarola, U.~Schollw\"ock, C.~Silva,
  B.~Surer, S.~Todo, S.~Trebst, M.~Troyer, M.~L. Wall, P.~Werner, S.~Wessel,
  \href{https://doi.org/10.1088/1742-5468/2011/05/p05001}{The {ALPS} project
  release 2.0: open source software for strongly correlated systems}, Journal
  of Statistical Mechanics: Theory and Experiment 2011~(05) (2011) P05001.
\newblock \href {https://doi.org/10.1088/1742-5468/2011/05/p05001}
  {\path{doi:10.1088/1742-5468/2011/05/p05001}}.
\newline\urlprefix\url{https://doi.org/10.1088/1742-5468/2011/05/p05001}

\bibitem{Parcollet2015CPC}
O.~Parcollet, M.~Ferrero, T.~Ayral, H.~Hafermann, I.~Krivenko, L.~Messio,
  P.~Seth,
  \href{https://www.sciencedirect.com/science/article/pii/S0010465515001666}{Triqs:
  A toolbox for research on interacting quantum systems}, Computer Physics
  Communications 196 (2015) 398--415.
\newblock \href {https://doi.org/https://doi.org/10.1016/j.cpc.2015.04.023}
  {\path{doi:https://doi.org/10.1016/j.cpc.2015.04.023}}.
\newline\urlprefix\url{https://www.sciencedirect.com/science/article/pii/S0010465515001666}

\bibitem{Gull2011RMP}
E.~Gull, A.~J. Millis, A.~I. Lichtenstein, A.~N. Rubtsov, M.~Troyer, P.~Werner,
  \href{https://link.aps.org/doi/10.1103/RevModPhys.83.349}{Continuous-time
  monte carlo methods for quantum impurity models}, Rev. Mod. Phys. 83 (2011)
  349--404.
\newblock \href {https://doi.org/10.1103/RevModPhys.83.349}
  {\path{doi:10.1103/RevModPhys.83.349}}.
\newline\urlprefix\url{https://link.aps.org/doi/10.1103/RevModPhys.83.349}

\bibitem{Rubtsov2005PRB}
A.~N. Rubtsov, V.~V. Savkin, A.~I. Lichtenstein,
  \href{https://link.aps.org/doi/10.1103/PhysRevB.72.035122}{Continuous-time
  quantum monte carlo method for fermions}, Phys. Rev. B 72 (2005) 035122.
\newblock \href {https://doi.org/10.1103/PhysRevB.72.035122}
  {\path{doi:10.1103/PhysRevB.72.035122}}.
\newline\urlprefix\url{https://link.aps.org/doi/10.1103/PhysRevB.72.035122}

\bibitem{Haule2007PRB}
K.~Haule, \href{http://link.aps.org/abstract/PRB/v75/e155113}{Quantum monte
  carlo impurity solver for cluster dynamical mean-field theory and electronic
  structure calculations with adjustable cluster base}, Phys. Rev. B 75~(15)
  (2007) 155113.
\newblock \href {https://doi.org/10.1103/PhysRevB.75.155113}
  {\path{doi:10.1103/PhysRevB.75.155113}}.
\newline\urlprefix\url{http://link.aps.org/abstract/PRB/v75/e155113}

\bibitem{Seth2016CPC}
P.~Seth, I.~Krivenko, M.~Ferrero, O.~Parcollet,
  \href{https://www.sciencedirect.com/science/article/pii/S001046551500404X}{Triqs/cthyb:
  A continuous-time quantum monte carlo hybridisation expansion solver for
  quantum impurity problems}, Computer Physics Communications 200 (2016)
  274--284.
\newblock \href {https://doi.org/https://doi.org/10.1016/j.cpc.2015.10.023}
  {\path{doi:https://doi.org/10.1016/j.cpc.2015.10.023}}.
\newline\urlprefix\url{https://www.sciencedirect.com/science/article/pii/S001046551500404X}

\bibitem{Wallerberger2019CPC}
M.~Wallerberger, A.~Hausoel, P.~Gunacker, A.~Kowalski, N.~Parragh, F.~Goth,
  K.~Held, G.~Sangiovanni,
  \href{https://www.sciencedirect.com/science/article/pii/S0010465518303217}{w2dynamics:
  Local one- and two-particle quantities from dynamical mean field theory},
  Computer Physics Communications 235 (2019) 388--399.
\newblock \href {https://doi.org/https://doi.org/10.1016/j.cpc.2018.09.007}
  {\path{doi:https://doi.org/10.1016/j.cpc.2018.09.007}}.
\newline\urlprefix\url{https://www.sciencedirect.com/science/article/pii/S0010465518303217}

\bibitem{Zitko2009PRB}
R.~\ifmmode~\check{Z}\else \v{Z}\fi{}itko, T.~Pruschke,
  \href{https://link.aps.org/doi/10.1103/PhysRevB.79.085106}{Energy resolution
  and discretization artifacts in the numerical renormalization group}, Phys.
  Rev. B 79 (2009) 085106.
\newblock \href {https://doi.org/10.1103/PhysRevB.79.085106}
  {\path{doi:10.1103/PhysRevB.79.085106}}.
\newline\urlprefix\url{https://link.aps.org/doi/10.1103/PhysRevB.79.085106}

\bibitem{Bulla2001PRB}
R.~Bulla, T.~A. Costi, D.~Vollhardt, {Finite-temperature numerical
  renormalization group study of the Mott transition}, Phys. Rev. B 64~(4)
  (2001) 045103.
\newblock \href {https://doi.org/10.1103/PhysRevB.64.045103}
  {\path{doi:10.1103/PhysRevB.64.045103}}.

\bibitem{Bulla2008RMP}
R.~Bulla, T.~A. Costi, T.~Pruschke,
  \href{http://link.aps.org/abstract/RMP/v80/p395}{{Numerical renormalization
  group method for quantum impurity systems}}, Rev. Mod. Phys. 80~(2) (2008)
  395.
\newblock \href {https://doi.org/10.1103/RevModPhys.80.395}
  {\path{doi:10.1103/RevModPhys.80.395}}.
\newline\urlprefix\url{http://link.aps.org/abstract/RMP/v80/p395}

\bibitem{White1992PRL}
S.~R. White, {Density matrix formulation for quantum renormalization groups},
  Phys. Rev. Lett. 69~(19) (1992) 2863--2866.
\newblock \href {https://doi.org/10.1103/PhysRevLett.69.2863}
  {\path{doi:10.1103/PhysRevLett.69.2863}}.

\bibitem{Hallberg1995PRB}
K.~A. Hallberg, {Density-matrix algorithm for the calculation of dynamical
  properties of low-dimensional systems}, Phys. Rev. B 52~(14) (1995)
  R9827--R9830.
\newblock \href {https://doi.org/10.1103/PhysRevB.52.R9827}
  {\path{doi:10.1103/PhysRevB.52.R9827}}.

\bibitem{Garcia2004PRL}
D.~J. Garc\'{\i}a, K.~Hallberg, M.~J. Rozenberg,
  \href{https://link.aps.org/doi/10.1103/PhysRevLett.93.246403}{Dynamical mean
  field theory with the density matrix renormalization group}, Phys. Rev. Lett.
  93 (2004) 246403.
\newblock \href {https://doi.org/10.1103/PhysRevLett.93.246403}
  {\path{doi:10.1103/PhysRevLett.93.246403}}.
\newline\urlprefix\url{https://link.aps.org/doi/10.1103/PhysRevLett.93.246403}

\bibitem{Schollwock2005RMP}
U.~Schollwöck, \href{http://link.aps.org/abstract/RMP/v77/p259}{{The
  density-matrix renormalization group}}, Rev. Mod. Phys. 77~(1) (2005) 259.
\newblock \href {https://doi.org/10.1103/RevModPhys.77.259}
  {\path{doi:10.1103/RevModPhys.77.259}}.
\newline\urlprefix\url{http://link.aps.org/abstract/RMP/v77/p259}

\bibitem{Wolf2015PRX}
F.~A. Wolf, A.~Go, I.~P. McCulloch, A.~J. Millis, U.~Schollw\"ock,
  \href{https://link.aps.org/doi/10.1103/PhysRevX.5.041032}{Imaginary-time
  matrix product state impurity solver for dynamical mean-field theory}, Phys.
  Rev. X 5 (2015) 041032.
\newblock \href {https://doi.org/10.1103/PhysRevX.5.041032}
  {\path{doi:10.1103/PhysRevX.5.041032}}.
\newline\urlprefix\url{https://link.aps.org/doi/10.1103/PhysRevX.5.041032}

\bibitem{Caffarel1994PRL}
M.~Caffarel, W.~Krauth,
  \href{https://link.aps.org/doi/10.1103/PhysRevLett.72.1545}{Exact
  diagonalization approach to correlated fermions in infinite dimensions: Mott
  transition and superconductivity}, Phys. Rev. Lett. 72 (1994) 1545--1548.
\newblock \href {https://doi.org/10.1103/PhysRevLett.72.1545}
  {\path{doi:10.1103/PhysRevLett.72.1545}}.
\newline\urlprefix\url{https://link.aps.org/doi/10.1103/PhysRevLett.72.1545}

\bibitem{Dolfen2006}
A.~Dolfen,
  \href{https://www.fz-juelich.de/cae/servlet/contentblob/1040316/publicationFile/23068/Dolfen.pdf}{Massively
  parallel exact diagonalization of strongly correlated systems}, Master's
  thesis, Forschungszentrum Juelich (2006).
\newline\urlprefix\url{https://www.fz-juelich.de/cae/servlet/contentblob/1040316/publicationFile/23068/Dolfen.pdf}

\bibitem{Perroni2007PRB}
C.~A. Perroni, H.~Ishida, A.~Liebsch,
  \href{https://link.aps.org/doi/10.1103/PhysRevB.75.045125}{Exact
  diagonalization dynamical mean-field theory for multiband materials: Effect
  of coulomb correlations on the fermi surface of
  ${\mathrm{na}}_{0.3}{\mathrm{coo}}_{2}$}, Phys. Rev. B 75 (2007) 045125.
\newblock \href {https://doi.org/10.1103/PhysRevB.75.045125}
  {\path{doi:10.1103/PhysRevB.75.045125}}.
\newline\urlprefix\url{https://link.aps.org/doi/10.1103/PhysRevB.75.045125}

\bibitem{Capone2007PRB}
M.~Capone, L.~de' Medici, A.~Georges,
  \href{https://link.aps.org/doi/10.1103/PhysRevB.76.245116}{Solving the
  dynamical mean-field theory at very low temperatures using the lanczos exact
  diagonalization}, Phys. Rev. B 76 (2007) 245116.
\newblock \href {https://doi.org/10.1103/PhysRevB.76.245116}
  {\path{doi:10.1103/PhysRevB.76.245116}}.
\newline\urlprefix\url{https://link.aps.org/doi/10.1103/PhysRevB.76.245116}

\bibitem{Weber2012PRB}
C.~Weber, A.~Amaricci, M.~Capone, P.~B. Littlewood,
  \href{http://link.aps.org/doi/10.1103/PhysRevB.86.115136}{{Augmented hybrid
  exact-diagonalization solver for dynamical mean field theory}}, Phys. Rev. B
  86 (2012) 115136.
\newblock \href {https://doi.org/10.1103/PhysRevB.86.115136}
  {\path{doi:10.1103/PhysRevB.86.115136}}.
\newline\urlprefix\url{http://link.aps.org/doi/10.1103/PhysRevB.86.115136}

\bibitem{Lu2017TEPJST}
Y.~Lu, M.~W. Haverkort,
  \href{https://doi.org/10.1140/epjst/e2017-70042-4}{Exact diagonalization as
  an impurity solver in dynamical mean field theory}, The European Physical
  Journal Special Topics 226~(11) (2017) 2549--2564.
\newblock \href {https://doi.org/10.1140/epjst/e2017-70042-4}
  {\path{doi:10.1140/epjst/e2017-70042-4}}.
\newline\urlprefix\url{https://doi.org/10.1140/epjst/e2017-70042-4}

\bibitem{Weisse2008}
H.~Fehske, R.~Schneider, A.~Weisse (Eds.),
  \href{https://doi.org/10.1007/978-3-540-74686-7_18}{Exact Diagonalization
  Techniques}, Springer Berlin Heidelberg, Berlin, Heidelberg, 2008, pp.
  529--544.
\newline\urlprefix\url{https://doi.org/10.1007/978-3-540-74686-7_18}

\bibitem{Sandvik2010ACP}
A.~W. Sandvik,
  \href{https://aip.scitation.org/doi/abs/10.1063/1.3518900}{Computational
  studies of quantum spin systems}, AIP Conference Proceedings 1297~(1) (2010)
  135--338.
\newblock \href
  {http://arxiv.org/abs/https://aip.scitation.org/doi/pdf/10.1063/1.3518900}
  {\path{arXiv:https://aip.scitation.org/doi/pdf/10.1063/1.3518900}}, \href
  {https://doi.org/10.1063/1.3518900} {\path{doi:10.1063/1.3518900}}.
\newline\urlprefix\url{https://aip.scitation.org/doi/abs/10.1063/1.3518900}

\bibitem{Siro2012CPC}
T.~Siro, A.~Harju,
  \href{https://www.sciencedirect.com/science/article/pii/S0010465512001452}{Exact
  diagonalization of the hubbard model on graphics processing units}, Computer
  Physics Communications 183~(9) (2012) 1884--1889.
\newblock \href {https://doi.org/https://doi.org/10.1016/j.cpc.2012.04.006}
  {\path{doi:https://doi.org/10.1016/j.cpc.2012.04.006}}.
\newline\urlprefix\url{https://www.sciencedirect.com/science/article/pii/S0010465512001452}

\bibitem{Borstnik2014PC}
U.~Borštnik, J.~VandeVondele, V.~Weber, J.~Hutter,
  \href{https://www.sciencedirect.com/science/article/pii/S0167819114000428}{Sparse
  matrix multiplication: The distributed block-compressed sparse row library},
  Parallel Computing 40~(5) (2014) 47--58.
\newblock \href {https://doi.org/https://doi.org/10.1016/j.parco.2014.03.012}
  {\path{doi:https://doi.org/10.1016/j.parco.2014.03.012}}.
\newline\urlprefix\url{https://www.sciencedirect.com/science/article/pii/S0167819114000428}

\bibitem{Siro2016CPC}
T.~Siro, A.~Harju,
  \href{https://www.sciencedirect.com/science/article/pii/S0010465516302065}{Exact
  diagonalization of quantum lattice models on coprocessors}, Computer Physics
  Communications 207 (2016) 274--281.
\newblock \href {https://doi.org/https://doi.org/10.1016/j.cpc.2016.07.018}
  {\path{doi:https://doi.org/10.1016/j.cpc.2016.07.018}}.
\newline\urlprefix\url{https://www.sciencedirect.com/science/article/pii/S0010465516302065}

\bibitem{Sharma2015CPC}
M.~Sharma, M.~Ahsan,
  \href{https://www.sciencedirect.com/science/article/pii/S0010465515001137}{Organization
  of the hilbert space for exact diagonalization of hubbard model}, Computer
  Physics Communications 193 (2015) 19--29.
\newblock \href {https://doi.org/https://doi.org/10.1016/j.cpc.2015.03.014}
  {\path{doi:https://doi.org/10.1016/j.cpc.2015.03.014}}.
\newline\urlprefix\url{https://www.sciencedirect.com/science/article/pii/S0010465515001137}

\bibitem{Kotliar2006RMP}
G.~Kotliar, S.~Y. Savrasov, K.~Haule, {\textit{ et al.}}, {Electronic structure
  calculations with dynamical mean-field theory}, Rev. Mod. Phys. 78~(3) (2006)
  865.
\newblock \href {https://doi.org/10.1103/RevModPhys.78.865}
  {\path{doi:10.1103/RevModPhys.78.865}}.

\bibitem{Lanczos1950JRNBSB}
C.~Lanczos, {An iteration method for the solution of the eigenvalue problem of
  linear differential and integral operators}, J. Res. Natl. Bur. Stand. B 45
  (1950) 255--282.
\newblock \href {https://doi.org/10.6028/jres.045.026}
  {\path{doi:10.6028/jres.045.026}}.

\bibitem{ARNOLDI1951QOAM}
W.~E. ARNOLDI, \href{http://www.jstor.org/stable/43633863}{The principle of
  minimized iterations in the solution of the matrix eigenvalue problem},
  Quarterly of Applied Mathematics 9~(1) (1951) 17--29.
\newline\urlprefix\url{http://www.jstor.org/stable/43633863}

\bibitem{A.Krylov1931BDLDSDL}
A.Krylov, \href{http://mi.mathnet.ru/izv5215}{De la r\'esolution num\'erique de
  l'\'equation servant \`a d\'eterminer dans des questions de m\'ecanique
  appliqu\'ee les fr\'equences de petites oscillations des syst\`emes
  mat\'eriels}, Bulletin de l'Acad\'emie des Sciences de l'URSS 5~(5) (1931)
  491--539.
\newline\urlprefix\url{http://mi.mathnet.ru/izv5215}

\bibitem{Polizzi2009PRB}
E.~Polizzi,
  \href{https://link.aps.org/doi/10.1103/PhysRevB.79.115112}{Density-matrix-based
  algorithm for solving eigenvalue problems}, Phys. Rev. B 79 (2009) 115112.
\newblock \href {https://doi.org/10.1103/PhysRevB.79.115112}
  {\path{doi:10.1103/PhysRevB.79.115112}}.
\newline\urlprefix\url{https://link.aps.org/doi/10.1103/PhysRevB.79.115112}

\bibitem{Georges2013ACMP}
A.~Georges, L.~de' Medici, J.~Mravlje, { Strong Correlations from Hund's
  Coupling.}, Annu.~Rev. Condens. Matter Phys. 45 (2013) 137--178.

\bibitem{Lehoucq1998}
R.~Lehoucq, D.~Sorensen, C.~Yang,
  \href{https://books.google.it/books?id=iMUea23N\_CQC}{ARPACK Users' Guide:
  Solution of Large-scale Eigenvalue Problems with Implicitly Restarted Arnoldi
  Methods}, Software, Environments, Tools, Society for Industrial and Applied
  Mathematics, 1998.
\newline\urlprefix\url{https://books.google.it/books?id=iMUea23N\_CQC}

\bibitem{Maschhoff1996}
K.~J. "Maschhoff, D.~C. Sorensen, P{\_}arpack: An efficient portable large
  scale eigenvalue package for distributed memory parallel architectures, in:
  J.~"Wa{\'{s}}niewski, J.~Dongarra, K.~Madsen, D.~Olesen (Eds.), Applied
  Parallel Computing Industrial Computation and Optimization, Springer Berlin
  Heidelberg, Berlin, Heidelberg, 1996, pp. 478--486.

\bibitem{Koch2008PRB}
E.~Koch, G.~Sangiovanni, O.~Gunnarsson,
  \href{https://link.aps.org/doi/10.1103/PhysRevB.78.115102}{Sum rules and bath
  parametrization for quantum cluster theories}, Phys. Rev. B 78 (2008) 115102.
\newblock \href {https://doi.org/10.1103/PhysRevB.78.115102}
  {\path{doi:10.1103/PhysRevB.78.115102}}.
\newline\urlprefix\url{https://link.aps.org/doi/10.1103/PhysRevB.78.115102}

\bibitem{Taranto2012PRB}
C.~Taranto, G.~Sangiovanni, K.~Held, M.~Capone, A.~Georges, A.~Toschi,
  \href{https://link.aps.org/doi/10.1103/PhysRevB.85.085124}{Signature of
  antiferromagnetic long-range order in the optical spectrum of strongly
  correlated electron systems}, Phys. Rev. B 85 (2012) 085124.
\newblock \href {https://doi.org/10.1103/PhysRevB.85.085124}
  {\path{doi:10.1103/PhysRevB.85.085124}}.
\newline\urlprefix\url{https://link.aps.org/doi/10.1103/PhysRevB.85.085124}

\bibitem{NumRec77}
W.~H. Press, B.~P. Flannery, S.~A. Teukolsky, W.~T. Vetterling,
  \href{http://www.worldcat.org/isbn/052143064X}{Numerical Recipes in FORTRAN
  77: The Art of Scientific Computing}, 2nd Edition, Cambridge University
  Press, 1992.
\newline\urlprefix\url{http://www.worldcat.org/isbn/052143064X}

\bibitem{MPI2}
{Message Passing Interface Forum},
  \href{http://www.mpi-forum.org/docs/mpi-2.2/mpi22-report.pdf}{MPI: A
  Message-Passing Interface Standard, Version 2.2}, High Performance Computing
  Center Stuttgart (HLRS), 2009.
\newline\urlprefix\url{http://www.mpi-forum.org/docs/mpi-2.2/mpi22-report.pdf}

\end{thebibliography}

\end{document}